\documentclass[titlepage,12pt]{article}
\usepackage{amssymb,amsmath,amsfonts}
\usepackage[utf8]{inputenc}
\usepackage{setspace}
\usepackage{axodraw4j}
\usepackage{epsfig}
\usepackage{graphicx}
\usepackage{color}
\usepackage{cite}
\usepackage{subfigure}
\usepackage{caption}
\usepackage{enumitem}

\makeatletter
\def\@sect#1#2#3#4#5#6[#7]#8{\ifnum #2>\c@secnumdepth
  \def\@svsec{}\else
  \refstepcounter{#1}\edef\@svsec{\csname the#1\endcsname.\hskip0.5em}\fi
  \@tempskipa #5\relax
  \ifdim \@tempskipa>\z@
    \begingroup
      #6\relax
      \@hangfrom{\hskip #3\relax\@svsec}{\interlinepenalty \@M #8\par}%
    \endgroup
    \csname #1mark\endcsname{#7}\addcontentsline
      {toc}{#1}{\ifnum #2>\c@secnumdepth \else
        \protect\numberline{\csname the#1\endcsname}\fi #7}%
  \else
    \def\@svsechd{#6\hskip #3\@svsec #8\csname #1mark\endcsname
      {#7}\addcontentsline{toc}{#1}{\ifnum #2>\c@secnumdepth \else
        \protect\numberline{\csname the#1\endcsname}\fi #7}}%
  \fi \@xsect{#5}}

\allowdisplaybreaks[4]

\newcommand{\R}{\text{Re}\,}
\newcommand{\as}{\alpha_s}
\newcommand{\QQbar}{Q{\bar Q}}
\newcommand{\mss}[1]{ {\mbox{\scriptsize #1}} }
\newcommand{\cN}{\ensuremath{\mathcal{N}}}
\newcommand{\cM}{\ensuremath{\mathcal{M}}}
\newcommand{\cA}{\ensuremath{\mathcal{A}}}
\newcommand{\cB}{\ensuremath{\mathcal{B}}}
\newcommand{\cO}{\ensuremath{\mathcal{O}}}
\begin{document}
\begin{titlepage}
  \begin{flushright}
MITP/14-066 \\
TTK-14-14\\
    \end{flushright}
\vspace{0.01cm}
\begin{center}
{\LARGE {\bf The real-virtual antenna functions for 
\boldmath$S\rightarrow Q\bar{Q} X$ at NNLO QCD} \\

\vspace{1.5cm}
\large{{\bf Oliver Dekkers\,$^{a,}$}\footnote{\tt dekkers@uni-mainz.de} and {\bf
Werner Bernreuther\,$^{b,}$}\footnote{\tt breuther@physik.rwth-aachen.de}}
\par\vspace{1cm}
{\small $^a$PRISMA Cluster of Excellence and Institut f\"ur Physik, \\
  Johannes Gutenberg-Universit\"at,
  D-55099 Mainz, Germany\\
$^b$Institut f\"ur Theoretische Physik, 
RWTH Aachen University, 52056 Aachen, Germany}\\
\par\vspace{1cm}
{\bf Abstract}\\
\parbox[t]{\textwidth}
{\small{ 
We determine, in the 
 antenna subtraction framework for handling infrared divergences in higher order QCD calculations, 
  the real-virtual antenna functions for processes involving the 
production of a pair of massive quarks  by an uncolored initial 
state at NNLO QCD.  The  integrated leading and subleading color real-virtual
antenna functions  are computed analytically in terms of
 (cyclotomic) harmonic polylogarithms. 
 As a by-product and check  we compute $R_Q=\sigma(e^+e^-\to \gamma^*\to  Q\bar{Q}
 X)/\sigma(e^+e^-\to \gamma^*\to\mu^+\mu^-)$ and compare with existing
 results.  Our result for $R_Q$  is exact to  order $\alpha_s^2$.
}}
}
\end{center}
\vspace*{0.7cm}

PACS number(s): 12.38.Bx, 13.88.+e, 14.65.Ha
\end{titlepage}
\setcounter{footnote}{0}
\renewcommand{\thefootnote}{\arabic{footnote}}
\setcounter{page}{1}

\section{Introduction} 
\label{introduction}

In this paper we report, within the antenna subtraction framework
\cite{Kosower:1997zr,Kosower:2003bh,GehrmannDeRidder:2005cm,Currie:2013vh}, 
 on the calculation  of the real-virtual antenna functions for processes involving the 
production of a pair of massive quarks  by an uncolored initial 
state $S$ at next-to-next-to leading order (NNLO)  QCD,
\begin{equation}\label{SQQNNLO}
 S \to Q \, {\bar Q }\, + X \, ,
\end{equation}
where $S$ denotes, for example, an $e^+e^-$ pair or an uncolored boson.
Antenna subtraction is a method  for handling infrared (IR), i.e., soft and
collinear divergences in higher order QCD calculations. The general
features of the method at NNLO QCD 
 were  presented in \cite{GehrmannDeRidder:2005cm}. Applications at
 NNLO QCD involving massless partons include $e^+e^-\to 2 \, {\rm
   jets}, 3\, {\rm
   jets}$ \cite{GehrmannDeRidder:2004tv,Weinzierl:2006ij,GehrmannDeRidder:2007jk,GehrmannDeRidder:2007hr} and
 $p p\to$ di-jets \cite{Ridder:2013mf}.
 For QCD processes involving massive quarks the method  was worked
 out to NLO in \cite{GehrmannDeRidder:2009fz,Abelof:2011jv}.
 Partial results exist for NNLO QCD processes with colored initial
 states and massive quarks in the final state \cite{Abelof:2011ap,Abelof:2012he,Abelof:2014fza}.
 For processes of the type \eqref{SQQNNLO} the un-integrated and
 integrated NNLO real radiation subtraction terms for the $Q{\bar Q}q{\bar
   q}$ and  $Q{\bar Q} g g$ final states were determined in \cite{Bernreuther:2011jt}
 and in \cite{Bernreuther:2013uma}, respectively.  The NNLO
 real-virtual antenna functions for  \eqref{SQQNNLO}, which were
 missing so far, are the subject of this paper.

At this point it seems appropriate to recall that other NNLO
subtraction techniques exist and have been successfully applied.
 A  method  was presented in
 \cite{Czakon:2010td,Czakon:2011ve,Czakon:2014oma} which can be used
 for massless and massive partons and was applied in the computation of the total hadronic $t\bar t$ cross
 section to order $\alpha_s^4$
 \cite{Baernreuther:2012ws,Czakon:2013goa}. 
 Other techniques for handling the IR divergences of the individual contributions to partonic
processes at NNLO QCD include the sector decomposition algorithm
\cite{Binoth:2000ps,Binoth:2003ak,Anastasiou:2003gr,Binoth:2004jv,Carter:2010hi}
 and the subtraction methods \cite{Weinzierl:2003fx,Frixione:2004is,Catani:2007vq,Somogyi:2006db,Somogyi:2006da,Somogyi:2013yk}.
 Very recently, a NNLO QCD generalization of the phase-space slicing method has been
 presented by  \cite{vonManteuffel:2014mva,Gao:2014nva} for $e^+e^-\to
 \gamma^*\to Q{\bar Q} X$.

Coming back to the issue of this paper, we outline in the next
section the construction of  separately
infrared (IR) finite contributions to  the NNLO differential cross section of
\eqref{SQQNNLO} from the four-parton, three-parton, and two-parton 
 final states within the antenna subtraction method. In Sec.~\ref{CompA13} we describe our
 computation of the integrated massive real-virtual antenna functions.
 As a first application
and check of our antenna subtraction terms we compute in
Sec.~\ref{sec:compR}  the contribution of order $\alpha_s^2$ to the
ratio $R_Q$ for inclusive massive quark-pair production by $e^+e^-$
annihilation via a virtual photon and compare with existing results
in the literature. We conclude in Sec.~\ref{sec:concl}.

 \section{The real-virtual subtraction term for reactions \eqref{SQQNNLO} }
\label{sec:revir}

The order $\as^2$ term $d\sigma_{\rm NNLO}$ in the strong-coupling expansion of
the differential cross section of \eqref{SQQNNLO}, $d\sigma=d\sigma_{\rm LO}+
d\sigma_{\rm NLO} + d\sigma_{\rm NNLO}$,  receives the following contributions:
i) the double virtual correction $d\sigma^{VV}_{\rm NNLO}$  associated
with the second-order matrix element of 
$S \to \QQbar$
(i.e., 2-loop times Born and 1-loop squared), 
 ii) the real-virtual cross section
 $d\sigma^{RV}_{\rm NNLO}$ associated with the second-order matrix element of 
$S \to \QQbar g$ (1-loop times Born),
iii) the double real contribution $d\sigma^{RR}_{\rm NNLO}$
associated with  the squared Born  amplitudes  
$S \to \QQbar gg$, 
$S \to \QQbar q{\bar q}$ 
(where  $q$ denotes a massless quark),
 and above the $4Q$ theshold,  
 $S \to \QQbar\QQbar.$ 
 The latter contribution is IR finite and is of no concern for the purpose of
this section. We will come back to it in Sec.~\ref{sec:compR}, where we
choose the initial state  in \eqref{SQQNNLO} to be a  virtual photon, $S=
\gamma^*$.

 All the terms given below denote renormalized quantities. The ultraviolet
divergences in the loop amplitudes are removed by on-shell renormalization of
the external quarks and gluons -- in the following, $m_Q$ denotes the on-shell mass of $Q$ --
 and by ${\overline{\rm MS}}$ renormalization of the strong coupling. 

The terms i), ii), iii) are, apart from the $\QQbar\QQbar$
contribution, separately IR divergent. 
Within the  subtraction method,  $d\sigma_{\rm NNLO}$
is given schematically by
\begin{eqnarray}
d\sigma_{\rm NNLO} = & \int_{\Phi_4}\left(d\sigma^{RR}_{\rm NNLO}  -d\sigma^{S}_{\rm NNLO}\right)
+  \int_{\Phi_3}\left(d\sigma^{RV}_{\rm NNLO}  -d\sigma^{T}_{\rm NNLO}\right) \nonumber \\
&  +  \int_{\Phi_2} d\sigma^{VV}_{\rm NNLO} +  \int_{\Phi_3} d\sigma^{T}_{\rm NNLO} + \int_{\Phi_4} d\sigma^{S}_{\rm NNLO}  \, .
\label{sub1NNLO}
\end{eqnarray}
The subscripts $\Phi_n$ denote $n$-particle phase-space integrals.
The integrands $d\sigma^{S}_{\rm NNLO}$ and $d\sigma^{T}_{\rm NNLO}$ denote 
the double-real subtraction terms (for $\QQbar q{\bar q}$ and  $\QQbar gg$) 
and the real-virtual subtraction term, respectively. The former are 
constructed such that the phase-space integrals
\begin{equation}  \label{sbdifcrx}
\int_{\Phi_4} \left[ 
  d \sigma^{RR, Q \bar{Q} gg}_\mss{NNLO} 
- d \sigma^{S, Q \bar{Q} gg}_\mss{NNLO}
\right]_{ \epsilon = 0 } \, , \qquad \sum_q \int_{\Phi_4} \left[ 
  d \sigma^{RR, Q \bar{Q} q{\bar q}}_\mss{NNLO} 
- d \sigma^{S, Q \bar{Q} q {\bar q}}_\mss{NNLO}
\right]_{ \epsilon = 0 } \, ,
\end{equation}
where $\epsilon =(4-d)/2$,  are finite in $d=4$ dimensions in all 
 single and double unresolved limits and  can be evaluated numerically.  
The real-virtual subtraction term
$d\sigma^{T}_{\rm NNLO}$ which is to be constructed such that 
 it reproduces both the implicit singularities 
in single unresolved phase space regions and the explicit poles 
of the real-virtual cross section 
$d\sigma^{RV}_{\rm NNLO}$,
 is the topic of this paper.
 The sum of the last three terms in \eqref{sub1NNLO} is also IR-finite, see
below. 

In order to make the cancellation of IR singularities explicit in
\eqref{sub1NNLO}, the integrals of these subtraction terms, denoted by 
$\int_{\Phi_3} d\sigma^{T}_{\rm NNLO}$ and  $\int_{\Phi_4}
 d\sigma^{S}_{\rm NNLO}$ in \eqref{sub1NNLO},
  must be computed over the phase-space regions where IR singularities
arise. In the antenna subtraction formalism, which is applied here,
 subtraction terms  are constructed from 
antenna functions and reduced matrix elements with remapped momenta. The antenna functions are
 universal building blocks  and  can be derived from the respective 
physical color-ordered squared matrix elements. 

\subsubsection*{Double real-radiation corrections}

The antenna subtraction terms 
$d \sigma^{S, Q \bar{Q} q \bar{q} }_\mss{NNLO}$ and 
$d \sigma^{S, Q \bar{Q} gg }_\mss{NNLO} $ for
the final states 
$\QQbar q{\bar q}$,  $\QQbar gg$ and their integrals
 over  appropriate phase space regions
 were computed in \cite{Bernreuther:2011jt} and \cite{Bernreuther:2013uma},
 respectively.  While some of these integrals, collectively denoted by  $\int_{\Phi_4}
d\sigma^{S}_{\rm NNLO}$ in \eqref{sub1NNLO}, contribute to the real-virtual
subtraction term and are therefore integrated over those regions of 
phase space which correspond to single-unresolved emission,
 others are combined with the double-virtual cross 
section and, hence, are integrated over the unresolved phase space of two partons. 
In order to keep track of these connections the subtraction terms are
decomposed into  a sum of three contributions \cite{Currie:2013vh}:
\begin{equation}\label{decsum2}
  \int_{\Phi_4} d \sigma^{S, Q \bar{Q} gg }_\mss{NNLO} 
  = \int_{\Phi_3} \int_1 d \sigma^{S,a, Q \bar{Q} g g}_\mss{NNLO}  
  + \int_{\Phi_2} \int_2 d\sigma^{S,b,2, Q \bar{Q} g g}_\mss{NNLO} 
  + \int_{\Phi_3} \int_1 d\sigma^{S,b,1, Q \bar{Q} g g}_\mss{NNLO} \,,
\end{equation}
where the integrands of the first two terms on the r.h.s.~cover, 
respectively, the singularities due to single and double
unresolved parton configurations in $d \sigma^{RR, Q \bar{Q} gg }_\mss{NNLO}$, 
whereas the third term removes spurious single unresolved limits from $d\sigma^{S,b,2, Q \bar{Q} g g}_\mss{NNLO}$. 
 A splitting  analogous to  \eqref{decsum2} is made for the subtraction term
$d\sigma^{S, Q \bar{Q} q{\bar q}}_\mss{NNLO}$.
Throughout this section the notation  $\int_n$ indicates the
analytic integration over the phase space of $n$ unresolved partons in 
$d\neq 4$ dimensions.

\subsubsection*{Real-virtual corrections}

Now, we move to the order $\alpha_s^2$ contribution of
\begin{equation} \label{S2QQg}
 S(q) \to Q( p_1) \, \bar{Q}( p_2 )\, g( p_3 )
\end{equation}
to eq.~\eqref{sub1NNLO} and construct the appropriate antenna subtraction terms 
following the lines of \cite{GehrmannDeRidder:2004tv,
GehrmannDeRidder:2005cm,Weinzierl:2006ij,Currie:2013vh}.
The real-virtual contribution to the differential cross section associated with the process \eqref{S2QQg} reads
\begin{eqnarray}
d \sigma^{RV,Q \bar{Q} g}_\mss{NNLO} & = & 
\cN_0
\left( 4 \pi \alpha_s \right)^{2} 
 C( \epsilon ) 
\left( N_c^2 - 1\right) 
d \Phi_3 ( p_1, p_2, p_3 ; q ) \,
 J^{(3)}_2 \! \left(  p_1, p_2, p_3 \right)
	\nonumber \\[0.5ex]
	& & {} \times \left[
	N_c \,	\delta \cM^{\mss{lc}}_{(3,1)} 
	- \frac{1}{N_c} \, \delta \cM^{\mss{sc}}_{(3,1)}  
	+ n_f \, \delta \cM^{\mss{f}}_{(3,1)} 
	+\delta \cM^{\mss{F}}_{(3,1)} \right] \, ,
\label{xs::QQg}
\end{eqnarray}
where $N_c$ denotes the number of colors, $n_f$ the number of massless
quarks, and  $\bar{C}(\epsilon) = 8 \pi^2 C( \epsilon ) = ( 4 \pi )^\epsilon 
e^{-\epsilon \gamma_E }$. The normalization factor $\cN_0$ comprises 
all non-QCD couplings as
well as the spin averaging factor for the initial state and the flux factor.
Furthermore, we have introduced a shorthand notation for the interference of
the tree-level and renormalized 1-loop amplitudes of \eqref{S2QQg}
(with all couplings and color matrices factored out):
\begin{equation} \label{delMX31}
\delta \cM^{X}_{(3,1)}
= 2\,\mbox{Re} \left[\cM^{*}_{(3,0)}( p_1, p_2, p_3) 
\cM^{X}_{(3,1)}(p_1, p_2, p_3) \right] .
\end{equation}
Here the superscript $X$ denotes the leading color ($lc$), subleading 
color ($sc$), and the massless ($f$) and massive ($F$) fermion contributions.
For the sake of brevity, we have dropped the dependence on the initial
state momenta.
 In \eqref{xs::QQg} and in the formulae below summation over all helicity states
is implicit. 

The measurement function $J^{(n)}_2$ in \eqref{xs::QQg} ensures that
only configurations are taken into account where $n$ outgoing partons form two
heavy quark jets. We emphasize that the following discussion 
applies to any infrared safe observable associated with the reactions \eqref{SQQNNLO}.

The renormalized one-loop amplitudes $\cM^{X}_{(3,1)}$ contain explicit IR
poles due to the exchange of virtual massless partons. On the other hand,
integrating \eqref{xs::QQg} over regions of the 3-particle phase space which
correspond to soft gluon emission leads to additional infrared singularities.
Both types of singularities have to be subtracted with appropriate counterterms
in order to perform the integration over the three-parton phase space
numerically. 

It is a well-known fact from NLO QCD calculations that the explicit IR poles in
\eqref{xs::QQg} can be removed by adding the 
subtraction terms $d\sigma^{S,a, Q \bar{Q} gg}_\mss{NNLO}$ and
 $d\sigma^{S,a, Q \bar{Q} q{\bar q}}_\mss{NNLO}$, integrated over the
one-unresolved parton antenna phase space, which govern the
singularities due to single unresolved emission in the squared tree-level
matrix elements of $S \to Q\bar{Q} gg$ and $S \to Q \bar{Q} q \bar{q}$:
\begin{equation}\label{dsigTa}
d \sigma^{T, a, Q \bar{Q} g}_\mss{NNLO} = 
- \int_1 d \sigma^{S,a, Q \bar{Q} g g}_\mss{NNLO}
- \sum_q \int_1 d \sigma^{S,a, Q \bar{Q} q \bar{q} }_\mss{NNLO} \, .
\end{equation}
 The terms on the r.h.s. are given  in
\cite{Bernreuther:2011jt,Bernreuther:2013uma}. 
\subsubsection*{One-loop single-unresolved subtraction term}
The singular behavior of \eqref{xs::QQg} in the limit where the external gluon
becomes soft requires another subtraction term which has the following
structure \cite{GehrmannDeRidder:2004tv,Currie:2013vh}:
\begin{eqnarray}
 d \sigma^{T,b,Q \bar{Q} g}_\mss{NNLO} & = & 
\cN_0
\left( 4 \pi \alpha_s \right)^{2} 
 C( \epsilon ) 
\left( N_c^2 - 1\right) 
d \Phi_3 ( p_1, p_2, p_3 ; q ) \,
J^{(2)}_2 \!  \left( \widetilde{p_{13}} , \widetilde{p_{23}}  \right) 
\nonumber \\
& & {} \times \bigg(
N_c \left[ A^1_3  \, \big|\cM_{(2,0)} \big|^2 
+ A^0_3 \, \delta \cM_{(2,1)} \right]
- \frac{1}{N_c} \left[ 
\tilde{A}^1_3 \, \big| \cM_{(2,0)} \big|^2 
+ A^0_3 \, \delta \cM_{(2,1)}  \right]
\nonumber \\
& & { } 
+ \left( n_f \, \hat{A}^1_{3,f} + \hat{A}^1_{3,F}\right)
\big| \cM_{(2,0)} \big|^2  \bigg), \quad 
\label{QQg::sub::b}
\end{eqnarray}
where 
\begin{equation} \label{delMX21}
\delta \cM_{(2,1)} 
= 2 \mbox{Re} \left[\cM^{*}_{(2,0)}( \widetilde{p_{13}}, \widetilde{p_{23}} )
\cM_{(2,1)} (\widetilde{p_{13}} , \widetilde{p_{23}} ) \right] \, ,
\end{equation}
and $\cM_{(2,i)}$, $i=0,1$, denote the tree-level and 1-loop matrix elements
 of $S \to Q {\bar Q}$ (with all couplings and color factors
stripped off). 
 These reduced amplitudes are evaluated at redefined on-shell momenta 
$\widetilde{p_{13}}, \widetilde{p_{23}}$ (i.e.~$\widetilde{p_{13}}^2 =
\widetilde{p_{23}}^2 = m_Q^2$) which are defined by Lorentz-invariant mappings
$p_1,p_2, p_3\to \widetilde{p_{13}}, \widetilde{p_{23}}$. These mappings are
given in \cite{Abelof:2011jv,Abelof:2011ap}.

\begin{figure}
  \begin{center}
    \subfigure[]{
     \includegraphics[scale=.2]{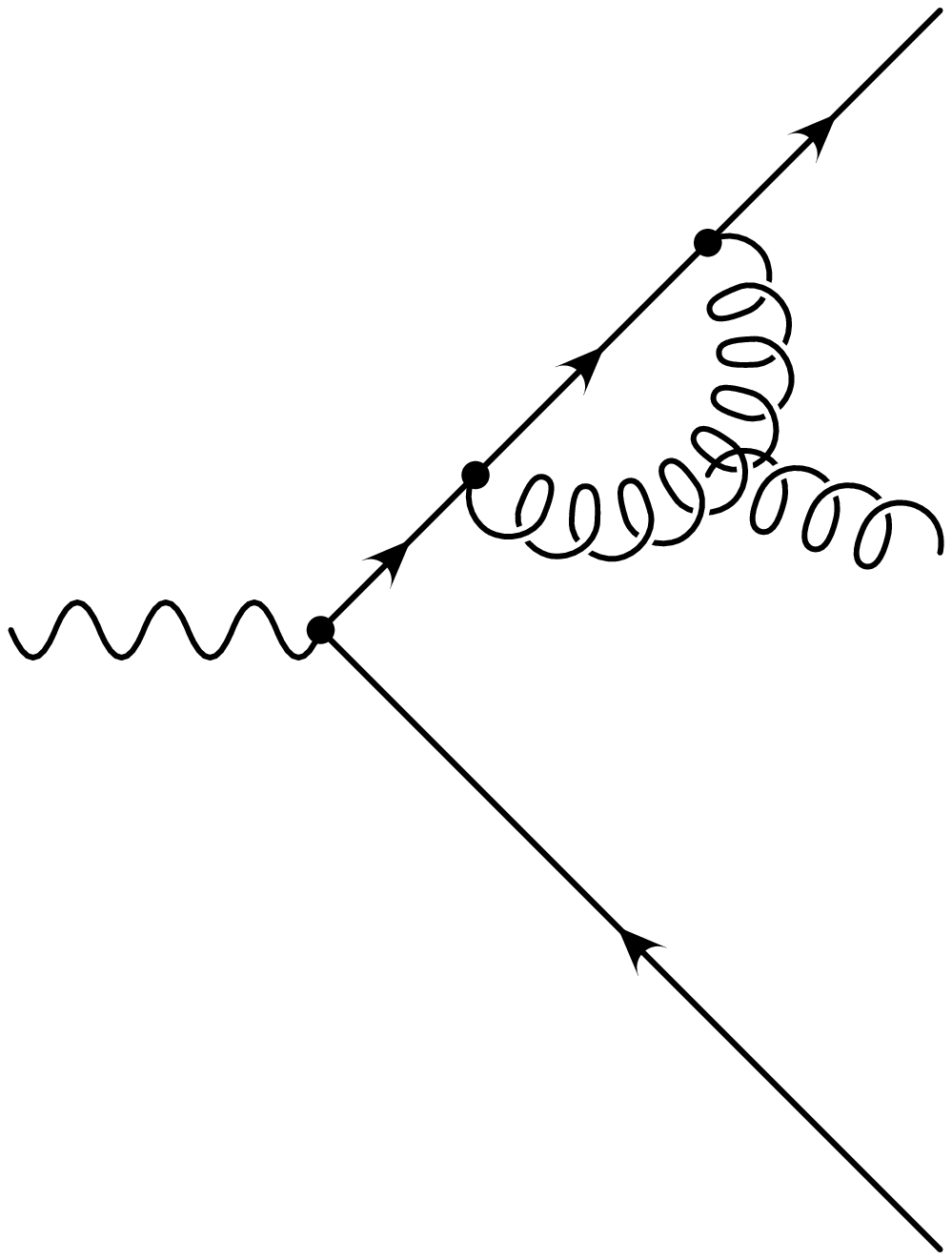}
    }
    \subfigure[]{
\includegraphics[scale=.2]{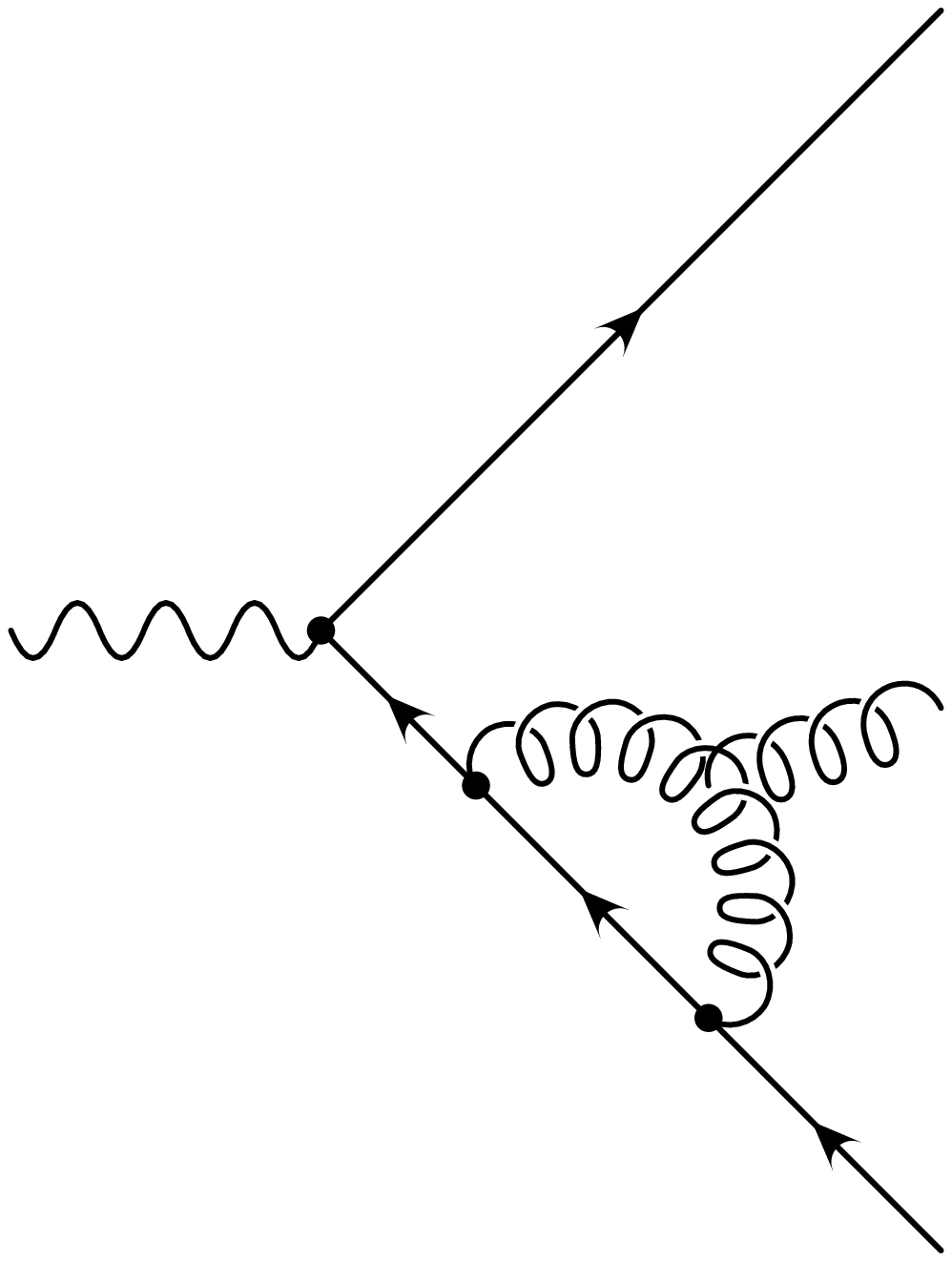}
    }
    \subfigure[]{
\includegraphics[scale=.2]{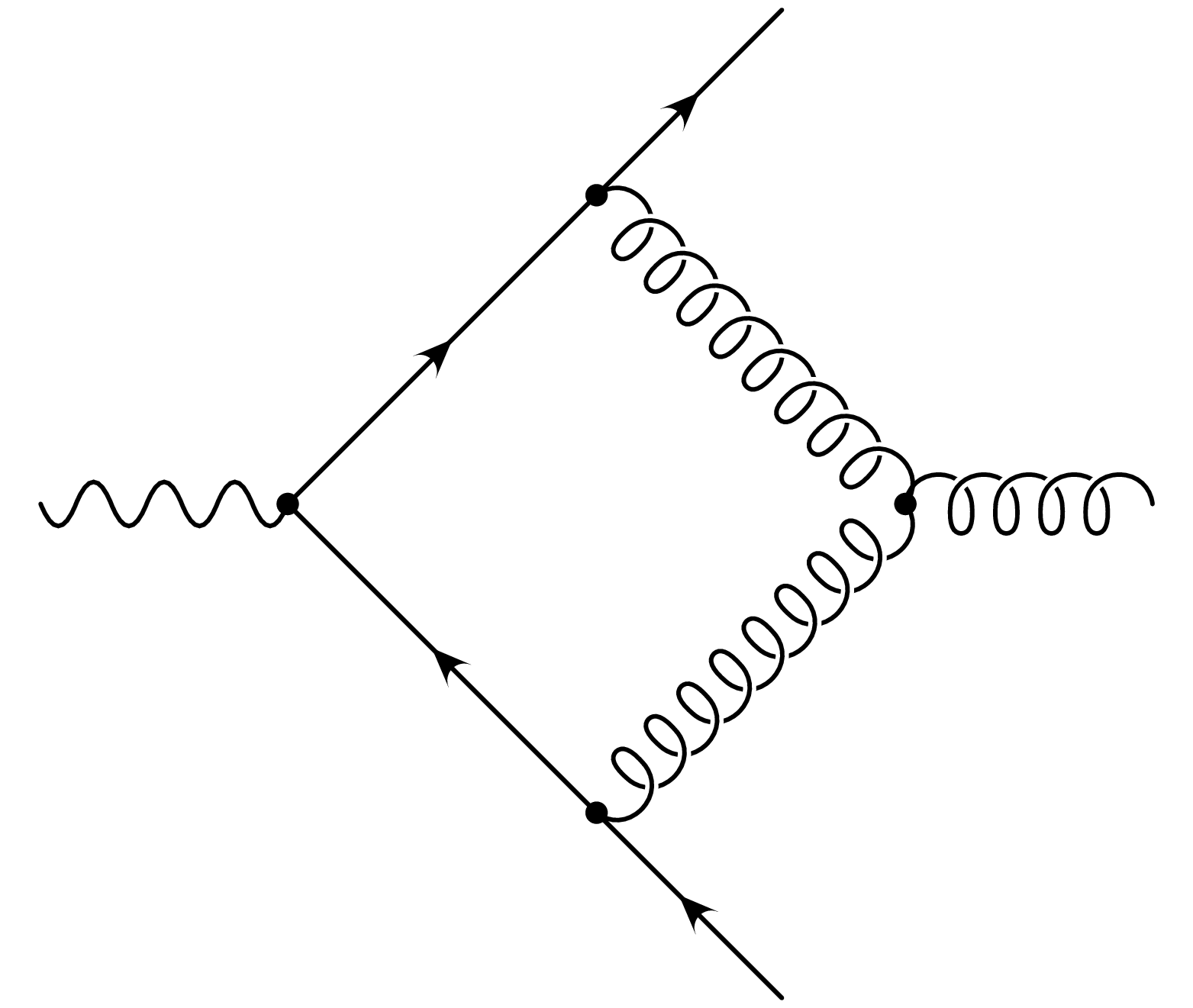}
    }
    \\
    \subfigure[]{
\includegraphics[scale=.2]{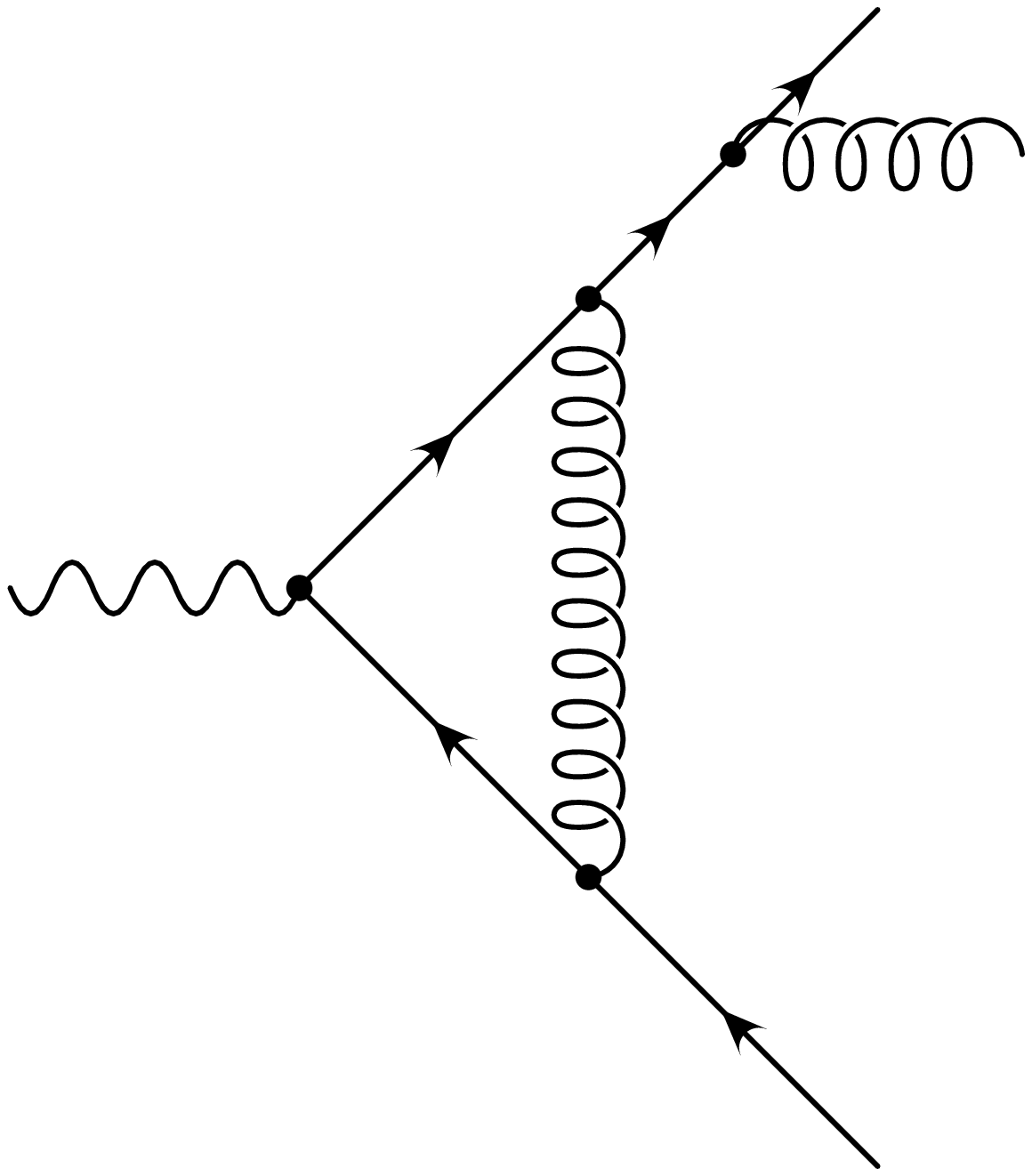}
    }
    \subfigure[]{
\includegraphics[scale=.2]{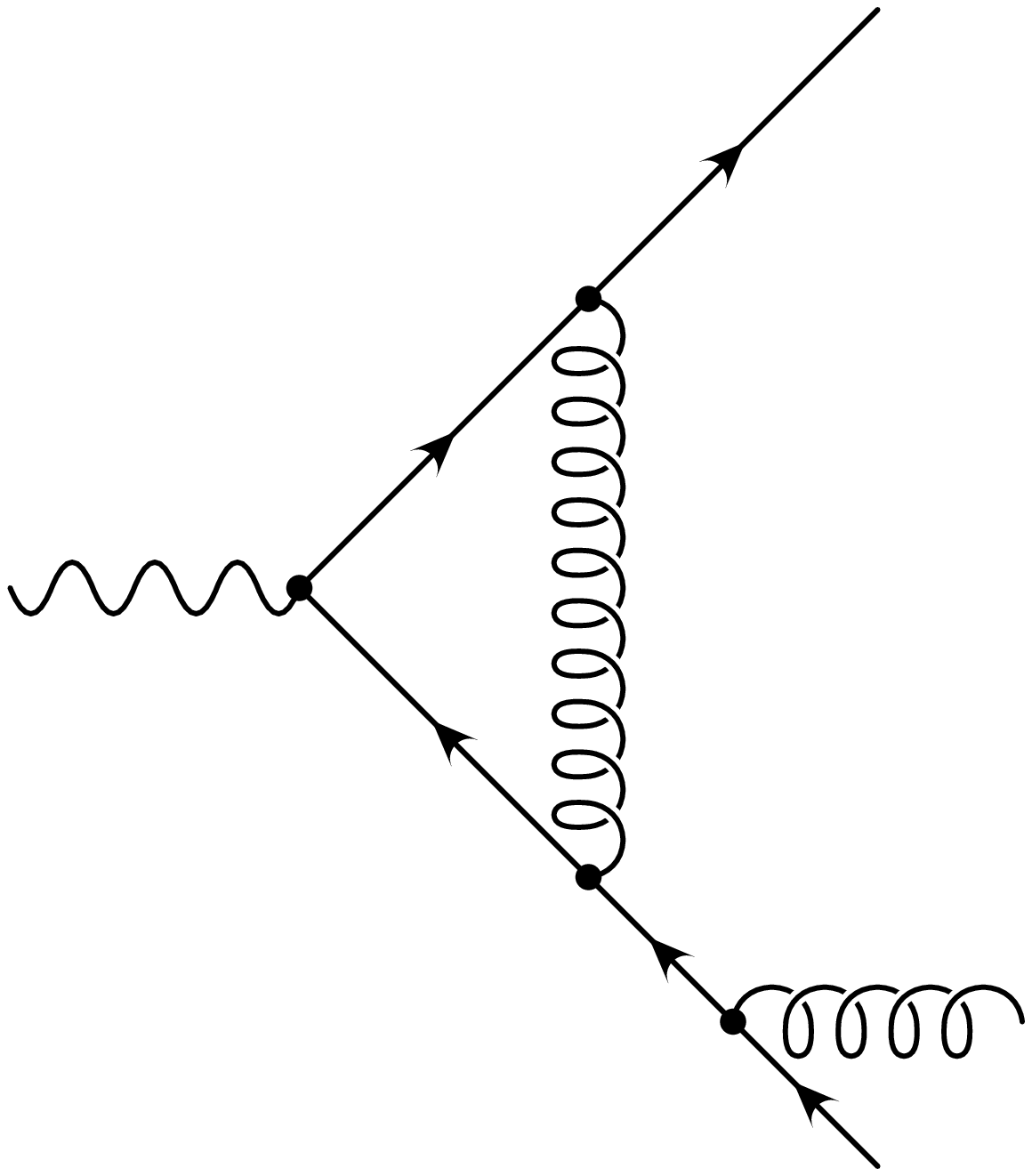}
    }
    \subfigure[]{
\includegraphics[scale=.2]{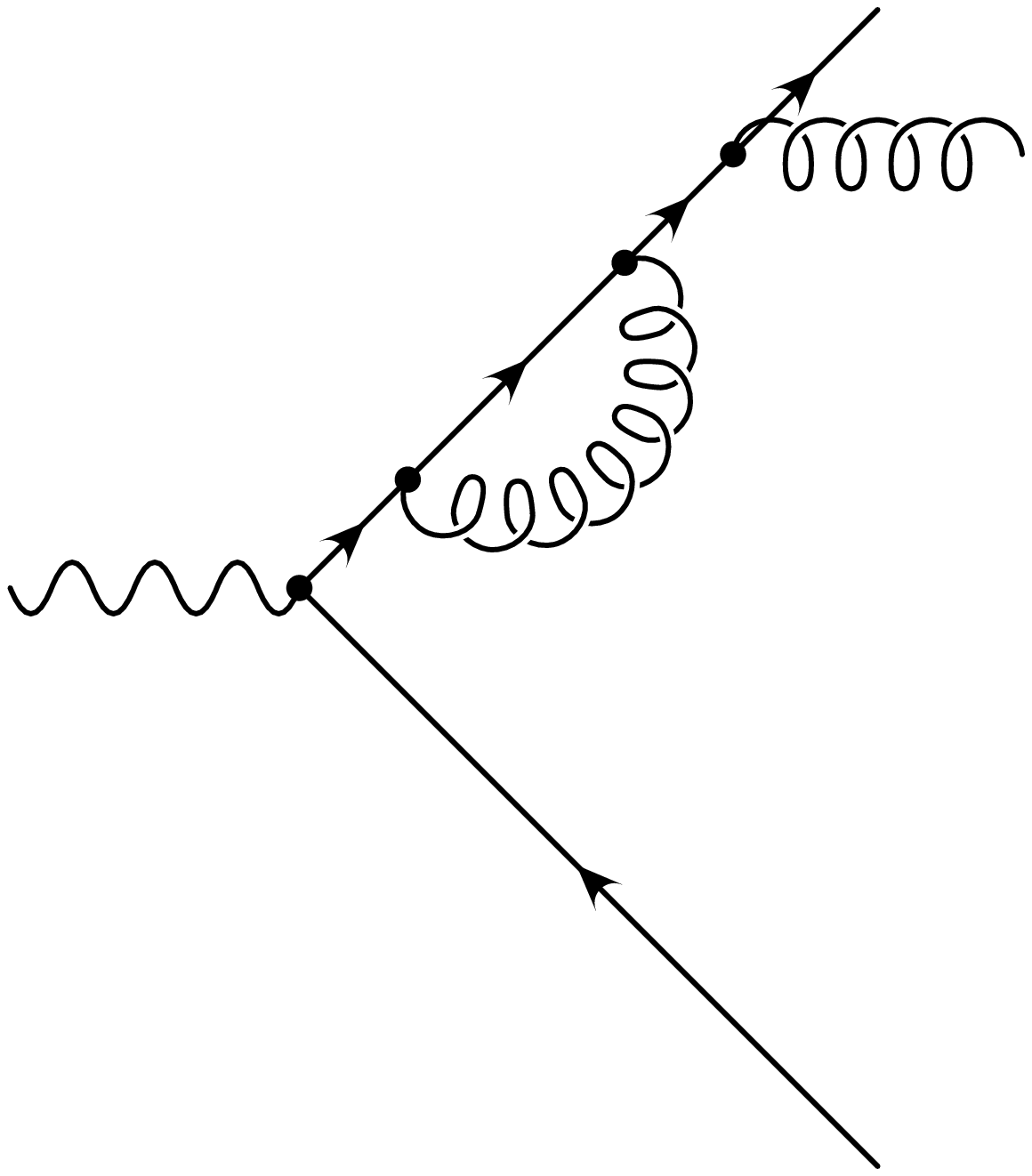}
    }
    \subfigure[]{
\includegraphics[scale=.2]{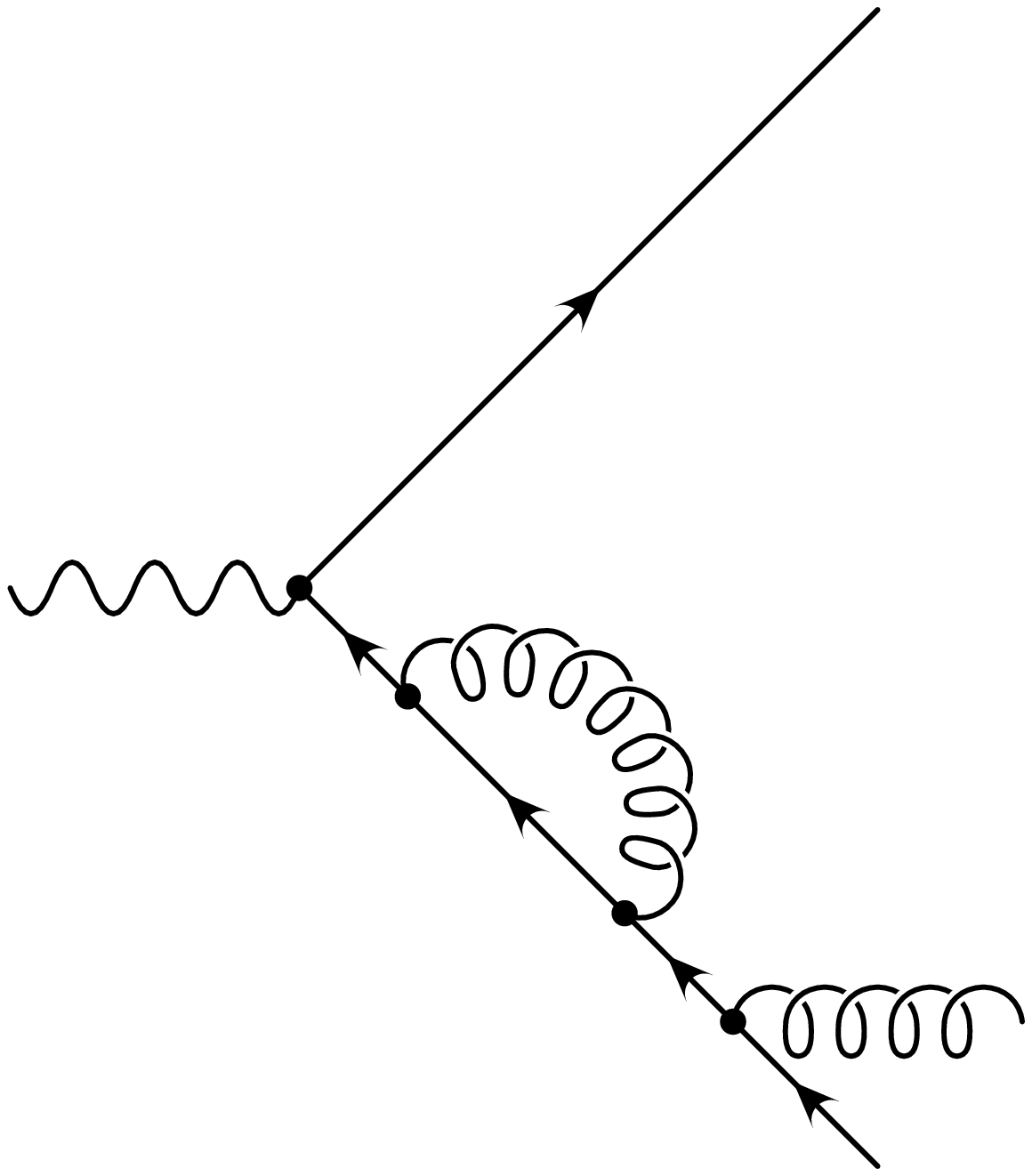}
    }
    \\
    \subfigure[]{
\includegraphics[scale=.2]{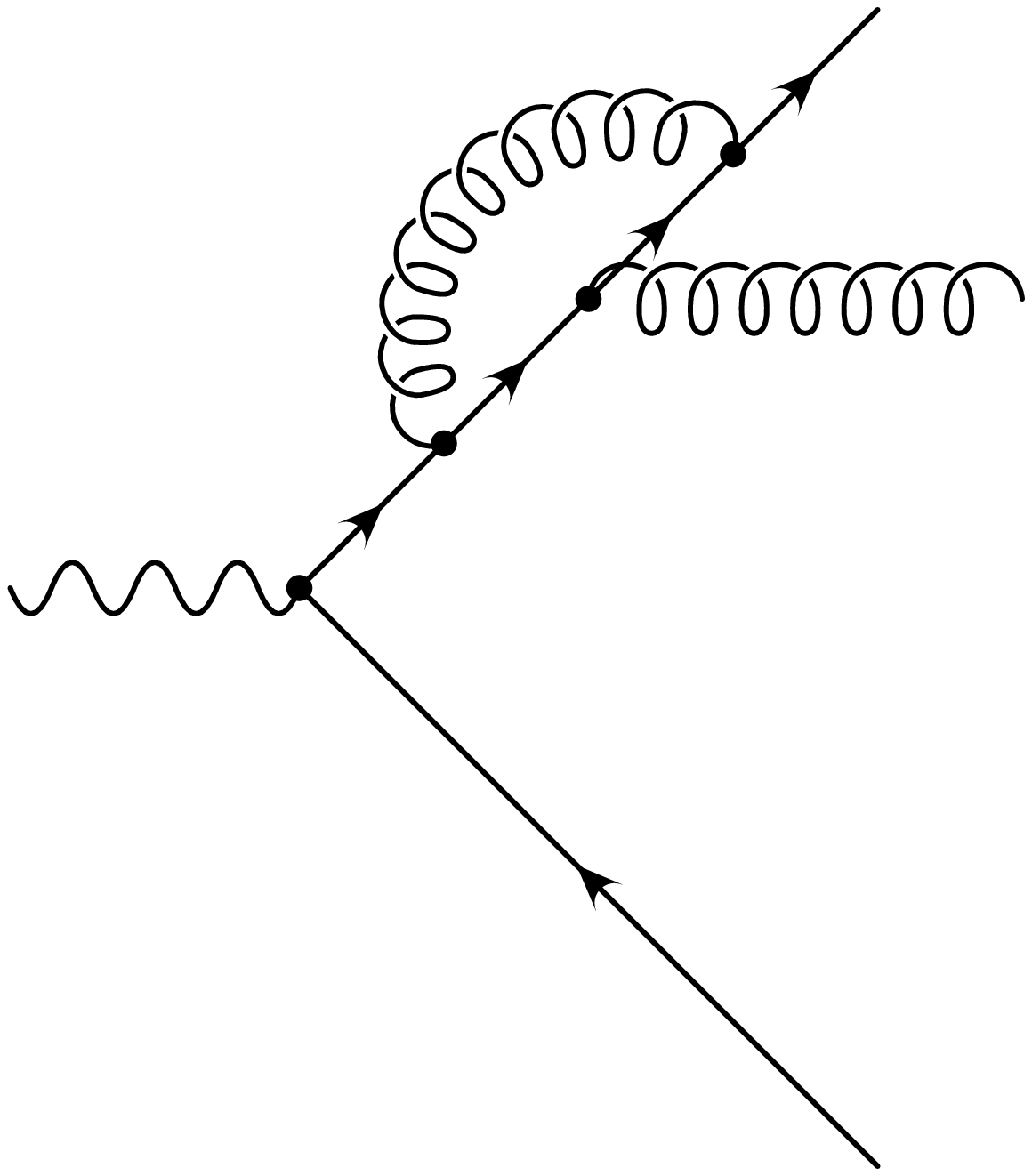}
    }
    \subfigure[]{
\includegraphics[scale=.2]{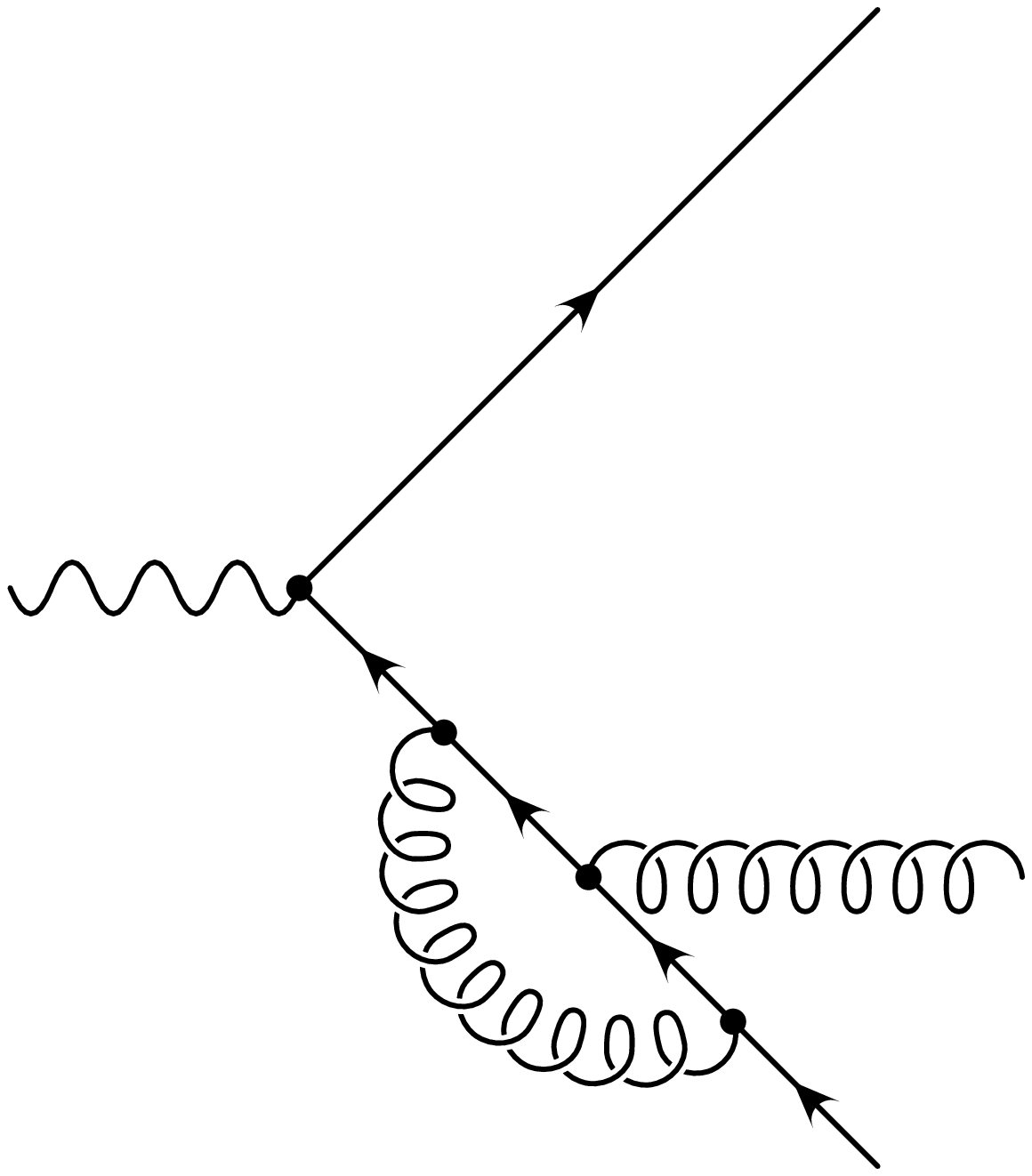}
    }
    \subfigure[]{
\includegraphics[scale=.2]{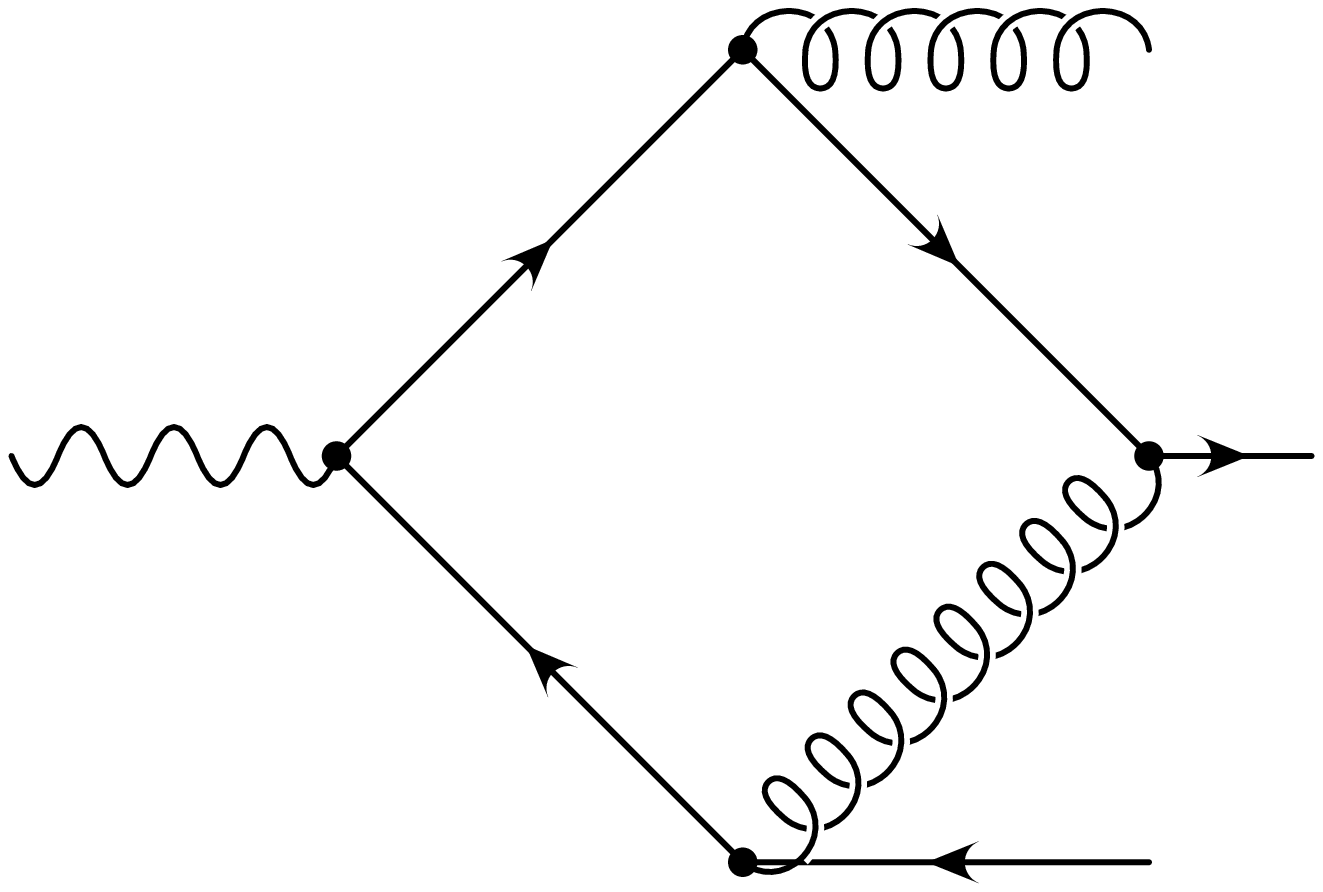}
    }
    \subfigure[]{
\includegraphics[scale=.2]{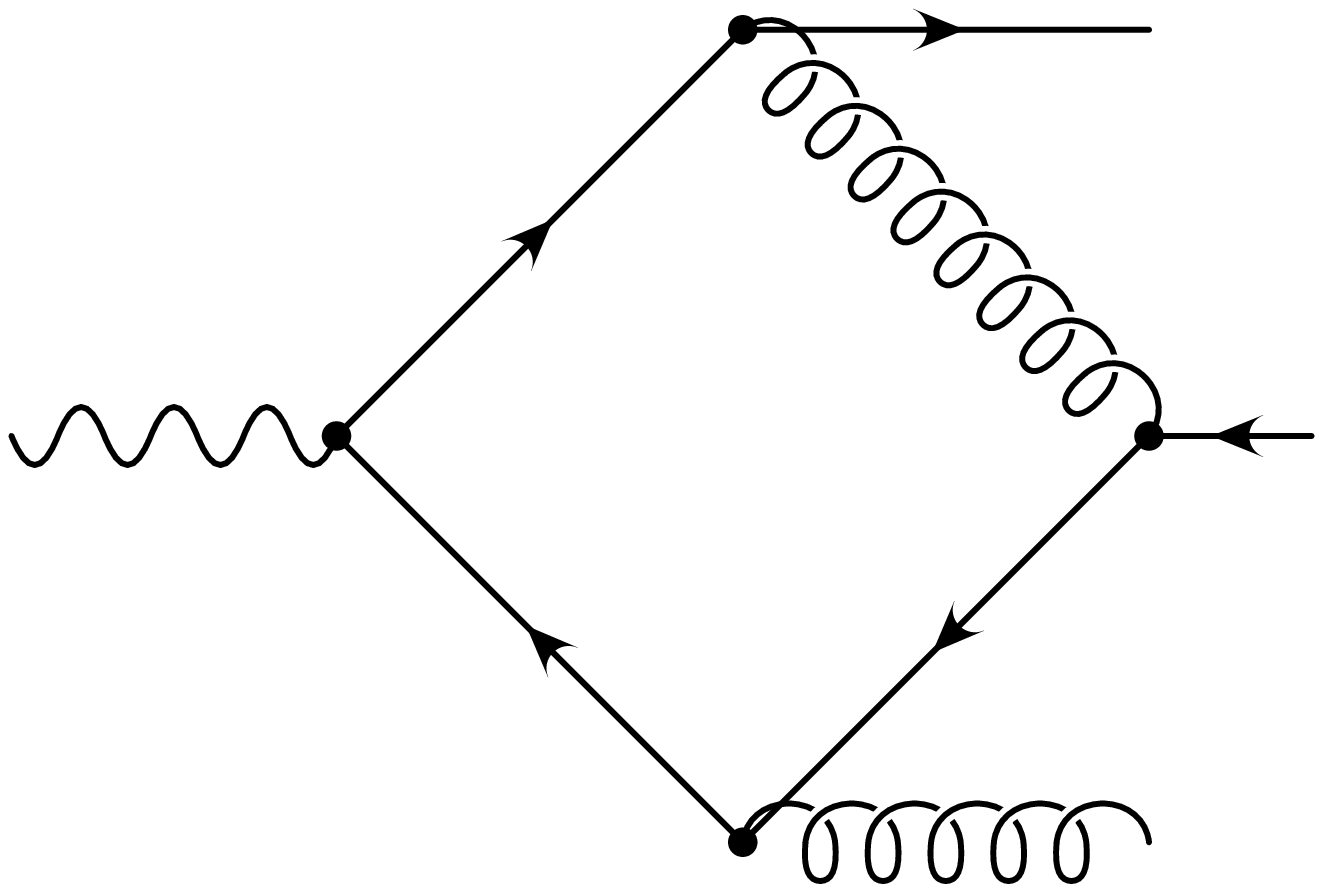}
    }
    %
    %
    %
    \caption{Diagrams that contribute to the one-loop QCD amplitude of
    $\gamma^*\to Q{\bar Q}g$. The ultraviolet counterterm diagrams are not
    shown. The interference of the tree-level diagram  with Figs. a) - c) 
    yields leading color contributions to eq. \eqref{xs::QQg}, the interference
    with Figs. d) -  g)  contains both leading and subleading color
    contributions, and the  interference  with   Figs.  h) - k) yields
    subleading color contributions.}
    \label{fig::A13}
  \end{center}
\end{figure}

 The massive tree-level antenna function $A^0_3$ and the massive 
one-loop antennae $A^1_{3}$, $\tilde{A}^1_{3}$,
$\hat{A}^1_{3,f}$, and $\hat{A}^1_{3,F}$ in \eqref{QQg::sub::b} depend 
 only on the original momenta $p_1, p_2, p_3$. 
 They can be derived once and for all from the tree-level and 1-loop 
matrix element of $\gamma^\ast \to Q \bar{Q} g$.
The corresponding 1-loop amplitudes are shown in Fig.~\ref{fig::A13}.
 We define
\begin{equation}
  A^1_3 \! \left( p_1, p_3, p_2 \right)
  = 
  \frac{ \delta \cM^{\gamma^\ast,\mss{lc}}_{(3,1)}  }
	{\big|\cM^{\gamma^\ast}_{(2,0)} \big|^2 }  
  - A^0_3 \! \left( p_1, p_3, p_2 \right)
  \frac{ \delta \cM^{\gamma^\ast}_{(2,1)} }
  {\big|\cM^{\gamma^\ast}_{(2,0)} \big|^2 }  ,
  \label{def::A13}
\end{equation}
and likewise for $\tilde{A}^1_{3}$ with $\delta
\cM^{\gamma^\ast,\mss{lc}}_{(3,1)}$ replaced by $\delta
\cM^{\gamma^\ast,\mss{sc}}_{(3,1)}$. %
The superscript $\gamma^\ast$ indicates that the quantities defined in 
\eqref{delMX31} and \eqref{delMX21} have to be determined specifically for 
$S = \gamma^\ast$. 
We note that $\cM^{\gamma^\ast}_{(2,0)}$ and  
  $\cM^{\gamma^\ast}_{(2,1)} $ only depend on $q^2$ 
  (and $m_Q$).

 The antenna functions 
 $\hat{A}^1_{3,f} = \delta \cM^{\gamma^\ast,f}_{(3,1)} / \big|\cM^{\gamma^\ast}_{(2,0)} \big|^2 $ and  
 $\hat{A}^1_{3,F} = \delta \cM^{\gamma^\ast,F}_{(3,1)} / \big|\cM^{\gamma^\ast}_{(2,0)} \big|^2 $ arise from 
 ultraviolet counterterm  contributions to the renormalized one-loop
 term \eqref{xs::QQg} and are therefore  proportional to the
 tree-level antenna function  $A^0_3$,
 which  was derived in \cite{GehrmannDeRidder:2009fz}.

The computation of $ \delta \cM^{\gamma^\ast, X}_{(3,1)}$,
$ \delta \cM^{\gamma^\ast}_{(2,1)}$, and 
$ \big|\cM^{\gamma^\ast}_{(2,0)} \big|^2$ is standard.
The resulting expressions for $A^1_{3}$, $\tilde{A}^1_{3}$,
$\hat{A}^1_{3,f}$, and $\hat{A}^1_{3,F}$ 
are quite long and will be presented elsewhere  \cite{DABB2014}.

 Because \eqref{QQg::sub::b} is not induced by an integrated double-real subtraction
term, it has to be added back to the double virtual cross section in
its integrated form.
 The integrated subtraction term can be cast into the form
\begin{eqnarray}
\int_1 d \sigma^{T,b,Q \bar{Q} g}_\mss{NNLO} & = & 
\cN_0
\left( 4 \pi \alpha_s \right)^{2} 
\left( C( \epsilon ) \right)^2
\left( N_c^2 - 1\right) 
d \Phi_2\!\left( p_1, p_2 ; q \right)
 J^{(2)}_2 \!  \left( p_1 , p_2 \right)
\nonumber \\[-0.5ex]
& & {} \times \bigg(
N_c \left[  \cA^1_3  \, \big|\cM_{(2,0)} \big|^2 
+ \cA^0_3 \, \delta \cM_{(2,1)} \right]
- \frac{1}{N_c} \left[ 
\tilde{\cA}^1_3 \, \big| \cM_{(2,0)} \big|^2 
+ \cA^0_3 \, \delta \cM_{(2,1)}  \right]
\nonumber \\
& & { } 
+ \left( n_f \, \hat{\cA}^1_{3,f} + \hat{\cA}^1_{3,F}\right)
\big| \cM_{(2,0)} \big|^2  \bigg), \qquad 
\label{rv::sub::b::int}
\end{eqnarray}
where we have introduced the integrated antenna functions
\begin{equation}
\cA^1_{3} \! \left( \epsilon, \mu^2/q^2 ; y \right)
 = 
\frac{1}{ C( \epsilon ) }
\int\! d\Phi_{X_{Qg\bar{Q}}} \, 
A_{3}^{1}\!\left( p_1,p_2,p_3\right)
 = 
\frac{1}{ C( \epsilon )\,P_{Q\bar{Q}} }
\int\! d\Phi_3 (p_1, p_2, p_3; q) \, 
A_{3}^{1}\!\left( p_1,p_2,p_3\right),
\label{sec::int_antennae::A13_def}
\end{equation}
and likewise for $\tilde{A}^1_{3}$, $\hat{A}^1_{3,f}$, $\hat{A}^1_{3,F}$,
and $A^0_{3}$.
In \eqref{sec::int_antennae::A13_def} we have made use of the fact that the
antenna phase-space measure $d\Phi_{X_{Qg\bar{Q}}}$ is defined as the
ratio of the usual three-particle phase-space $d\Phi_3$ (associated with
a massive quark-antiquark pair and a gluon) and the phase-space volume
 $P_{Q\bar{Q}} \equiv P_{Q\bar{Q}} ( q^2, m_Q^2)$ of a heavy $Q \bar{Q}$ pair with total four-momentum $q$. 
Consequently, the integrated massive one-loop antenna functions
$\cA^1_{3}$, $\tilde{\cA}^1_{3}$, $\hat{\cA}^1_{3,f}$, and
$\hat{\cA}^1_{3,F}$ depend on $\mu^2/q^2$ (where $\mu$ is the
renormalization scale) and the dimensionless
variable $y=(1-\beta)/(1+\beta)$, where $\beta=\sqrt{1-4m_Q^2/q^2}$.
In \eqref{rv::sub::b::int} the squared tree-level matrix element 
$|\cM_{(2,0)}|^2$ and the interference term  $\delta \cM_{(2,1)}$
depend on the heavy (anti)quark momenta $p_1$ and $p_2$.

Because the  antenna functions $\hat{A}^1_{3,f}$  and  $\hat{A}^1_{3,F}$  
are  proportional to the massive tree-level antenna $A^0_3$, the
integrated antenna functions $\hat{\cA}^1_{3,f}$  and $\hat{\cA}^1_{3,F}$
are proportional to the integrated massive antenna function $\cA^0_3$,
 which was already determined in the context of NLO antenna subtraction
\cite{GehrmannDeRidder:2009fz,Abelof:2011jv}.
 The calculation of $\cA^1_{3}$ and $\tilde{\cA}^1_{3}$  will be described in the
next section. 

\subsubsection*{Compensation terms for oversubtracted poles}
 The 1-loop antenna functions contain explicit IR poles which arise
from the loop integrations in $\delta \cM^{\gamma^\ast,X}_{(3,1)}$ and $\delta
\cM^{\gamma^\ast}_{(2,1)}$. The same holds for the reduced 1-loop matrix
element $\cM_{(2,1)}$ in \eqref{QQg::sub::b}. These poles do not
coincide with respective poles of the real-virtual contribution 
\eqref{xs::QQg} to the differential cross section, because those have already been
subtracted with the help of $d \sigma^{T,a,Q \bar{Q} g}_\mss{NNLO}$. 
In order to remove these spurious singularities one has to introduce an
additional subtraction term which we denote by  $d \sigma^{T,c,Q \bar{Q}
g}_\mss{NNLO}$.
 This additional subtraction term is given by (cf.~eq.~\eqref{decsum2}
and \cite{GehrmannDeRidder:2004tv,Currie:2013vh})
\begin{equation} \label{dsigTc}
  d \sigma^{T,c,Q \bar{Q} g}_\mss{NNLO} = 
- \int_1 d \sigma^{S,b,1,Q\bar{Q} g g }_\mss{NNLO} 
- \sum_q \, \int_1 d \sigma^{S,b,1,Q\bar{Q} q \bar{q}}_\mss{NNLO} .
\end{equation}

To conclude, combining Eqs.~\eqref{xs::QQg},  \eqref{dsigTa},
\eqref{QQg::sub::b}, and \eqref{dsigTc}
yields an expression which is free of (explicit and implicit) singularities
in the entire three-parton phase space in $d = 4$ dimensions:
\begin{equation} \label{ph3RVfinite}
 \int_{\Phi_3} \left[ 
  d \sigma^{RV,Q \bar{Q} g}_\mss{NNLO}
- d \sigma^{T, a, Q \bar{Q} g }_\mss{NNLO} 
- d \sigma^{T, b, Q \bar{Q} g }_\mss{NNLO} 
- d \sigma^{T, c, Q \bar{Q} g }_\mss{NNLO} 
\right]_{ \epsilon = 0 }
= \mbox{finite}.
\end{equation}
As discussed above, the introduction of $d \sigma^{T, a, Q \bar{Q} g
}_\mss{NNLO}$ and $d \sigma^{T, c, Q \bar{Q} g }_\mss{NNLO}$ is counterbalanced
by integrated double-real subtraction terms. Hence, only $d \sigma^{T, b, Q
\bar{Q} g }_\mss{NNLO}$ has to be added back  in its integrated form \eqref{rv::sub::b::int}
 to the double virtual contribution.

\subsubsection*{Double-virtual corrections}
In order to cancel the explicit IR poles of  the double virtual contribution 
$d\sigma^{VV,Q \bar{Q} }_\mss{NNLO}$ to   \eqref{sub1NNLO}, we have to add the
integrated subtraction terms resulting from $d\sigma^{S,b,2, Q \bar{Q} g
g}_\mss{NNLO}$ and $d\sigma^{S,b,2, Q \bar{Q} q \bar{q} }_\mss{NNLO}$ 
(cf.~eq.~\eqref{decsum2})  and \eqref{rv::sub::b::int}: 
\begin{equation} \label{dsi2nnlosub}
 \int_{\Phi_2} \left[ 
  d \sigma^{VV,Q \bar{Q} }_\mss{NNLO}
+ \int_1 d \sigma^{T,b, Q \bar{Q} g }_\mss{NNLO} 
+ \int_2 d \sigma^{S,b,2, Q \bar{Q} g g}_\mss{NNLO}
+ \sum_q  \int_2 d \sigma^{S,b,2, Q \bar{Q} q \bar{q} }_\mss{NNLO} 
\right]_{ \epsilon = 0 }
= \mbox{finite}.
\end{equation}
The integrations over the respective phase spaces of the unresolved partons
have to be done in $d = 4 - 2 \epsilon$. Only after summing up all the terms in
\eqref{dsi2nnlosub} and analytic cancellation of the IR poles, one can take
the limit $\epsilon \to 0$ and perform the remaining integration over the
two-parton phase space. 

Using \eqref{rv::sub::b::int} the subtraction term for the double-virtual
correction has the following structure:
\begin{eqnarray}
\lefteqn{
\int_1 d \sigma^{T,b, Q \bar{Q} g }_\mss{NNLO} 
+ \int_2 d \sigma^{S,b,2, Q \bar{Q} g g}_\mss{NNLO}
+ \sum_q \int_2 d \sigma^{S,b,2, Q \bar{Q} q \bar{q} }_\mss{NNLO} 
} \quad 
\nonumber \\
& = & 
\cN_0
\left( \bar{C} \! \left( \epsilon \right) \right)^2
\left( \frac{ \alpha_s }{ 2 \pi } \right)^2
\left( N_c^2 - 1 \right)
 d \Phi_2\!\left( p_1, p_2 ; q \right)
 J^{(2)}_2 \!  \left( p_1 , p_2 \right)
\nonumber \\
& & {} \times \bigg\{ N_c \bigg[ \Big(
\cA^0_4  + \cA^1_3 \Big) \big|\cM_{(2,0)} \big|^2 
+ \cA^0_3 \, \delta \cM_{(2,1)} \bigg]
\nonumber \\
& & {} - \frac{1}{ N_c } \left[ \left(
\frac{1}{2} \,  \tilde{\cA}^0_4  + \tilde{\cA}^1_3 \right)
\big|\cM_{(2,0)} \big|^2 
+ \cA^0_3 \, \delta \cM_{(2,1)} \right]
\nonumber \\
& & {} + \left[ n_f \left( \cB^0_4  + \hat{\cA}^1_{3,f} \right) 
+ \hat{\cA}^1_{3,F} \right]  \big|\cM_{(2,0)} \big|^2 \bigg\} \, .
\label{VV::sub}
\end{eqnarray}
We recall that the integrated antenna functions
 $\cA^0_4, \tilde{\cA}^0_{4},  \cB^0_{4},$
$\cA^1_{3}$, $\tilde{\cA}^1_{3}$, $\hat{\cA}^1_{3,f}$, $\hat{\cA}^1_{3,F}$, 
and  $\cA^0_3$  depend on $\mu^2/q^2$  and  $y=(1-\beta)/(1+\beta)$.  
The massive tree-level quark-antiquark antenna $\cA^0_3$ was computed in 
\cite{GehrmannDeRidder:2009fz,Abelof:2011jv}, and the integrated massive
four-parton antenna functions $\cA^0_4, \tilde{\cA}^0_{4},$ and  $\cB^0_{4}$
which result from the second and third integrals in the first line of
\eqref{VV::sub}, were determined in \cite{Bernreuther:2013uma} and
\cite{Bernreuther:2011jt}, respectively. Finally, explicit results for
the integrated massive 1-loop antenna $\cA^1_{3}$, $\tilde{\cA}^1_{3}$,
$\hat{\cA}^1_{3,f}$, and $\hat{\cA}^1_{3,F}$ are presented in the next section.
 
In conclusion, the expressions  \eqref{sbdifcrx},  \eqref{ph3RVfinite},  and
\eqref{dsi2nnlosub} constitute, within the antenna subtraction method,
the IR-finite four-parton, three-parton, and two-parton contributions to
\eqref{sub1NNLO}.   Let us stress again that the antenna subtraction terms
that we have introduced in this section are applicable to any reaction of 
the type \eqref{SQQNNLO} for an arbitrary colorless initial state $S$.

\section{Computation of ${\cal A}^1_3$ and ${\tilde{\cal A}}^1_3$}
\label{CompA13}

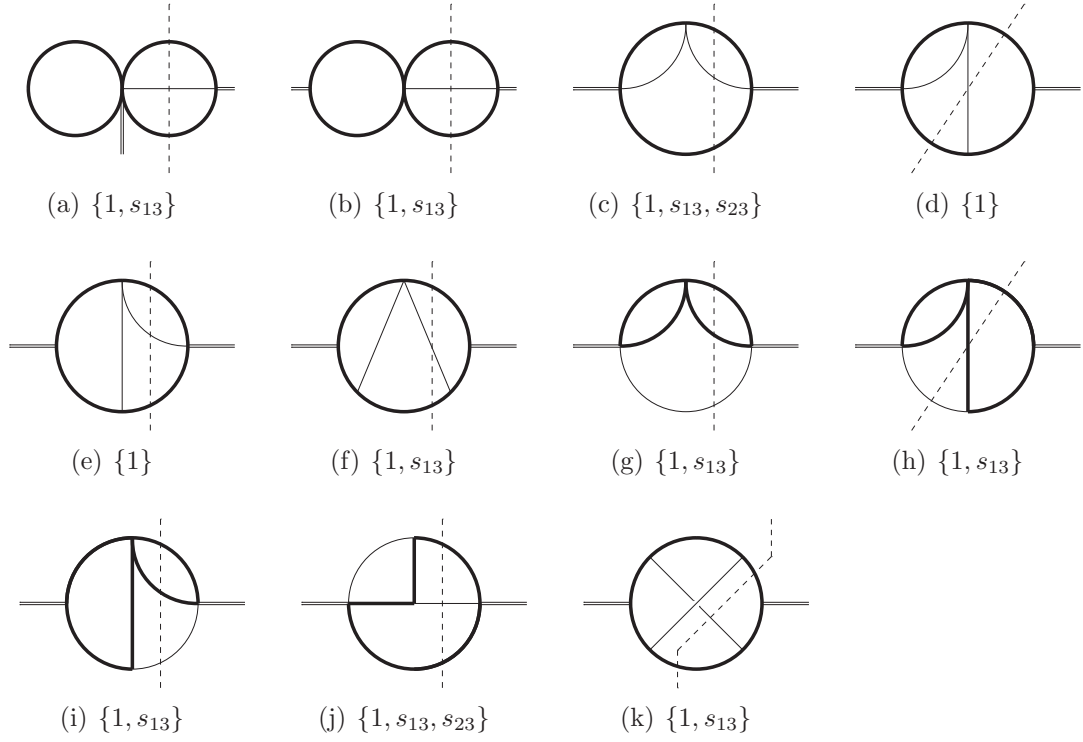
\begin{figure}
  \begin{center}
  %
  \subfigure[ $\{1, s_{13} \}$ ]{
  \resizebox{0.18\linewidth}{!}{
  \begin{picture}(386,290) (159,-127)
    \SetWidth{1.0}
    \SetColor{White}
    \Vertex(176,18){16}
    \SetColor{Black}
    \Line[double,sep=4](352,18)(352,-94)
    \Line[double,sep=4](512,18)(544,18)
    \Line[dash,dashsize=10](432,162)(432,-126)
    \SetWidth{6.0}
    \Arc(272,18)(80,180,540)
    \Arc(432,18)(80,180,540)
    \SetWidth{1.0}
    \Line(352,18)(512,18)
  \end{picture}
  }}
  \quad
  %
  \subfigure[ $\{1, s_{13} \}$ ]{
  \resizebox{0.18\linewidth}{!}{
  \begin{picture}(386,290) (159,-127)
    \SetWidth{1.0}
    \SetColor{Black}
    \Line[double,sep=4](160,18)(192,18)
    \Line[double,sep=4](512,18)(544,18)
    \Line[dash,dashsize=10](432,162)(432,-126)
    \SetWidth{6.0}
    \Arc(272,18)(80,180,540)
    \Arc(432,18)(80,180,540)
    \SetWidth{1.0}
    \Line(352,18)(512,18)
  \end{picture}
  }}
  \quad
  %
  \subfigure[ $\{1, s_{13}, s_{23} \}$ ]{
  \resizebox{0.18\linewidth}{!}{
  \begin{picture}(386,290) (159,-127)
      \SetWidth{6.0}
      \SetColor{Black}
      \Arc(352,18)(112,90,450)
      \SetWidth{1.0}
      \Line[double,sep=4](160,18)(240,18)
      \Line[double,sep=4](464,18)(544,18)
      \Arc(242.667,127.333)(109.366,-91.397,1.397)
      \Arc[clock](461.333,127.333)(109.366,-88.603,-181.397)
      \Line[dash,dashsize=10](400,162)(400,-126)
    \end{picture}
  }} 
  \quad
  %
  \subfigure[ $\{ 1 \}$ ]{
  \resizebox{0.18\linewidth}{!}{
  \begin{picture}(386,290) (159,-127)
    \SetWidth{6.0}
    \SetColor{Black}
    \Arc(352,18)(112,90,450)
    \SetWidth{1.0}
    \Line[double,sep=4](160,18)(240,18)
    \Line[double,sep=4](464,18)(544,18)
    \Arc(242.667,127.333)(109.366,-91.397,1.397)
    \Line(352,130)(352,-94)
    \Line[dash,dashsize=10](448,162)(256,-126)
  \end{picture}
  }}
  \\[2ex]
  %
  \subfigure[ $\{ 1 \}$ ]{
  \resizebox{0.18\linewidth}{!}{
  \begin{picture}(386,290) (159,-127)
    \SetWidth{6.0}
    \SetColor{Black}
    \Arc(352,18)(112,90,450)
    \SetWidth{1.0}
    \Line[double,sep=4](160,18)(240,18)
    \Line[double,sep=4](464,18)(544,18)
    \Arc[clock](461.333,127.333)(109.366,-88.603,-181.397)
    \Line[dash,dashsize=10](400,162)(400,-126)
    \Line(352,130)(352,-94)
  \end{picture}
  }}
  \quad
  %
  \subfigure[ $\{ 1, s_{13} \}$ ]{
  \resizebox{0.18\linewidth}{!}{
  \begin{picture}(386,290) (159,-127)
    \SetWidth{6.0}
    \SetColor{Black}
    \Arc(352,18)(112,90,450)
    \SetWidth{1.0}
    \Line[double,sep=4](160,18)(240,18)
    \Line[double,sep=4](464,18)(544,18)
    \Line[dash,dashsize=10](400,162)(400,-126)
    \Line(352,130)(432,-62)
    \Line(352,130)(272,-62)
  \end{picture}
  }} 
  \quad
  %
  \subfigure[ $\{1, s_{13} \}$ ]{
  \resizebox{0.18\linewidth}{!}{
  \begin{picture}(386,290) (159,-127)
    \SetWidth{1.0}
    \SetColor{Black}
    \Line[double,sep=4](160,18)(240,18)
    \Line[double,sep=4](464,18)(544,18)
    \SetWidth{6.0}
    \Arc(242.667,127.333)(109.366,-91.397,1.397)
    \Arc[clock](461.333,127.333)(109.366,-88.603,-181.397)
    \SetWidth{1.0}
    \Line[dash,dashsize=10](400,162)(400,-126)
    \Arc(352,18)(112,-180,0)
    \SetWidth{6.0}
    \Arc[clock](352,18)(112,-180,-360)
  \end{picture}
  }}
  \quad
  %
  \subfigure[ $\{1, s_{13} \}$ ]{
  \resizebox{0.18\linewidth}{!}{
  \begin{picture}(386,290) (159,-127)
    \SetWidth{1.0}
    \SetColor{Black}
    \Arc(352,18)(112,90,450)
    \Line[double,sep=4](160,18)(240,18)
    \Line[double,sep=4](464,18)(544,18)
    \SetWidth{6.0}
    \Arc[clock](242.667,127.333)(109.366,1.397,-91.397)
    \SetWidth{1.0}
    \Line[dash,dashsize=10](448,162)(256,-126)
    \SetWidth{6.0}
    \Line(352,130)(352,-94)
    \Arc(352,18)(112,-90,90)
    \Arc[clock](352,18)(112,-180,-360)
  \end{picture}
  }} 
  \\[2ex]
  %
  \subfigure[ $\{1, s_{13} \}$ ]{
  \resizebox{0.18\linewidth}{!}{
  \begin{picture}(386,290) (159,-127)
    \SetWidth{1.0}
    \SetColor{Black}
    \Arc(352,18)(112,90,450)
    \Line[double,sep=4](160,18)(240,18)
    \Line[double,sep=4](464,18)(544,18)
    \SetWidth{6.0}
    \Arc[clock](461.333,127.333)(109.366,-88.603,-181.397)
    \SetWidth{1.0}
    \Line[dash,dashsize=10](400,162)(400,-126)
    \SetWidth{6.0}
    \Line(352,130)(352,-94)
    \Arc[clock](352,18)(112,-90,-270)
    \Arc[clock](352,18)(112,-180,-360)
  \end{picture}
  }}
  \quad
  %
  \subfigure[ $\{1, s_{13}, s_{23} \}$ ]{
  \resizebox{0.18\linewidth}{!}{
  \begin{picture}(386,290) (159,-127)                                           
    \SetWidth{1.0}                                                              
    \SetColor{Black}                                                            
    \Line[double,sep=4](160,18)(240,18)                                         
    \Line[double,sep=4](464,18)(544,18)                                         
    \Line[dash,dashsize=10](400,162)(400,-126)
    \Arc[clock](352,18)(112,-90,-270)
    \SetWidth{6.0}
    \Arc[clock](352,18)(112,90,-90)
    \Line(352,130)(352,18)
    \Line(240,18)(352,18)
    \SetWidth{1.0}
    \Line(464,18)(352,18)
    \SetWidth{6.0}
    \Arc(352,18)(112,-180,0)
  \end{picture}
  }}
  \quad
  %
  \subfigure[ $\{1, s_{13} \}$ ]{
  \resizebox{0.18\linewidth}{!}{
  \begin{picture}(386,290) (159,-127)
    \SetWidth{1.0}
    \SetColor{Black}
    \Line[double,sep=4](160,18)(240,18)
    \Line[double,sep=4](464,18)(544,18)
    \SetWidth{6.0}
    \Arc(352,18)(112,90,450)
    \SetWidth{1.0}
    \Line(432,98)(272,-62)
    \Line(357,13)(432,-62)
    \Line(347,23)(272,98)
    \Line[dash,dashsize=10](320,-126)(320,-62)
    \Line[dash,dashsize=10](320,-62)(480,98)
    \Line[dash,dashsize=10](480,98)(480,162)
  \end{picture}
  }}
  \subfigure{ \resizebox{0.18\linewidth}{!}{ } }
  \caption{The 11 topologies which correspond to the 22 master
    integrals which determine
  the antenna  functions  ${\cal A}^1_3$ and ${\tilde{\cal A}}^1_3$. 
   Bold (thin) lines refer to massive (massless) scalar propagators.
  The dashed line represents the three-particle cut. The  external double line
  represents the external off-shell momentum $q$. The terms in the curly
  brackets below each topology denote additional factors in the corresponding
  integrand of the combined phase-space and loop integral.}
  \label{I_RV_Topos}
 \end{center}
\end{figure}

The integrated antenna functions ${\cal A}^1_3$ and  ${\tilde{\cal A}}^1_3$ are
computed as follows. We represent the d-dimensional three-particle phase-space
measure in terms of cut-propagators \cite{Anastasiou:2002yz}. We perform
an integration-by-parts reduction \cite{Chetyrkin:1981qh} using the computer
implementation ${\tt FIRE}$ \cite{Smirnov:2008iw} of the Laporta algorithm
\cite{Laporta:2001dd} in order to express  these two integrated antenna
functions by 22 master integrals  which correspond to three-particle cuts
through three-loop scalar self-energy type Feynman integrals, involving massive
and massless scalar propagators, associated with 11 different topologies, see
Fig.~\ref{I_RV_Topos}. The terms in the curly brackets listed below each
topology $(s_{ij}\equiv 2 p_i\cdot p_j)$ denote factors which multiply the
integrand of the corresponding integral. For instance, the two integrals which
are represented by the diagram Fig.~\ref{I_RV_Topos}k are, with
$A\in\{1,s_{13}\}$: 
\begin{equation} \label{MItopok}
I_{(k)}^A= \R\left[i^{-1}\int d\Phi_3^d
\frac{A}{s_{23}}\int\frac{d^dk}{(2\pi)^d}
\frac{1}{D(k,m,0)D(k,0,p_2)D(k,m,p_1+p_2)D(k,m,p_1+p_2+p_3)}\right] \, ,
\end{equation}
where $D(k,m,P)=(k-P)^2-m^2+i\eta$ and $d\Phi_3^d$ denotes the
3-particle phase-space measure in $d$ dimensions. 

The master integrals are computed by the method of differential equations
\cite{Kotikov:1990kg,Remiddi:1997ny,Gehrmann:1999as,Argeri:2007up}, i.e., we
derive inhomogeneous first order differential equations  in the variables $q^2$
and $y =(1-\beta)/(1+\beta)$ for
each master integral $I_{(\alpha)}^A$, $\alpha=a,...,k.$ The solutions
 $I_{(\alpha)}^A$ associated with the topologies (a), (b), (g), (h),
   (i), and (k) of  Fig.~\ref{I_RV_Topos}   can be expressed in terms of harmonic polylogarithms (HPL)
\cite{Remiddi:1999ew} with argument $y$ up to and including weight four.
In the solutions  $I_{(\alpha)}^A$ associated 
with the topologies (c), (d), (e), (f), and (j) of Fig.~\ref{I_RV_Topos} new
structures appear and these solutions can be expressed in terms of cyclotomic harmonic
polylogarithms \cite{Blumlein:2009ta,Ablinger:2010kw,Ablinger:2013hcp,Ablinger:2011te,Ablinger:2013cf}. The solutions of the  first-order
differential equations involve integration constants which we fix by
determining the boundary condition that  each master integral $I_{(\alpha)}^A$ must
satisfy at the $\QQbar$ production threshold $\beta=0$. For this purpose we
expand  the $d$-dimensional 1-loop integral and the phase-space measure which
make up each $I_{(\alpha)}^A$ im powers of $\epsilon$ and $\beta$. Phase-space
integration then yields the coefficients of the Laurent series in  $\epsilon$
of the  $I_{(\alpha)}^A$ at $y=1$. The integration constants which involve HPL at $y=1$
can be straightforwardly expressed in terms of transcendental numbers \cite{Remiddi:1999ew,Maitre:2005uu}. 
Those integration constants that involve cyclotomic HPL at $y=1$ can be
computed either by numerical evaluation of the respective integral
representation  of the  cyclotomic HPL or by (computer) algebraic reduction to
transcendental numbers
\cite{Blumlein:2009ta,Ablinger:2010kw,Ablinger:2013hcp,Ablinger:2011te,Ablinger:2013cf,AB2014}.
 In this way, the  22 master integrals  $I_{(\alpha)}^A$
and, with these integrals, the integrated antenna functions  are calculated
analytically. The complete expressions of the individual $I_{(\alpha)}^A$  will be given
elsewhere \cite{DABB2014}.

Below we list the IR-divergent pieces of the integrated
one-loop antenna functions defined in eq.~\eqref{sec::int_antennae::A13_def}.
 For the integrated leading and subleading color one-loop antennae, $\cA^1_{3}$
and $\tilde{\cA}^1_{3}$, we find:
\begin{eqnarray}
  \lefteqn{
  \cA^1_{3} \!\left( \epsilon, \mu^2/q^2 ;  y \right) 
  \; = \;
  \left( \frac{\mu^2}{q^2} \right)^{2 \epsilon}
  \Bigg\{ 
  %
  \frac{1}{ \epsilon^3 }
  \Bigg[ 
  -\frac{1}{2}
  + \frac{1}{2} \left(
  1
  -\frac{1}{1-y}
  -\frac{1}{1+y}
  \right)
  \text{H}( 0 ; y)
  \Bigg]
  } \quad 
  \nonumber \\
  & & {} 
  %
  %
  + \frac{1}{ \epsilon^2 }
  \Bigg[ 
  \left(
  - \frac{5}{3}
  + \frac{11}{12 (1-y)}
  - \frac{25}{12 (1+y)}
  + \frac{2 y+1}{2 \left(y^2+4 y+1\right)}
  \right) \text{H}(0 ; y)
  -4 \text{H}( 1 ; y)
  \nonumber \\
  & & {} 
  +\left(
  1
  -\frac{1}{1-y}
  -\frac{1}{1+y}
  \right)
  \left( 
  4 \text{H}(0,1;y)
  + 3 \text{H}(-1,0 ; y)
  - \text{H}(1,0 ; y)
  - \frac{7 \zeta(2) }{2}
  \right)
  \nonumber \\
  & & {} 
  + \left(
  1
  -\frac{1}{1-y}
  -\frac{1}{1+y}
  +\frac{1}{(1-y)^2}
  +\frac{1}{(1+y)^2}
  \right)
  \text{H}(0,0;y)
  -\frac{3 y}{y^2+4 y+1}
  -\frac{49}{12} 
  \Bigg]
  \nonumber \\
  & & {} 
  %
  %
  + \frac{1}{ \epsilon }
  \Bigg[
  \left(
    \frac{11}{6}
    -\frac{11}{6} \left(
      1
      - \frac{1}{1-y}
      - \frac{1}{1+y}
    \right)
    \text{H}(0,y)
  \right) 
  \ln \left( \frac{ \mu^2 }{ q^2 } \right)
  \nonumber \\
  & & {} 
  +\bigg(
  \frac{121+50 \pi ^2}{24 (y+1)}+\frac{1}{12}
\left(-254-25 \pi ^2\right)+\frac{-457-50 \pi ^2}{24 (y-1)}-\frac{7 \pi ^2}{6
(y-1)^2}-\frac{7 \pi ^2}{6 (y+1)^2}
  \nonumber \\
  & & {} 
  +\frac{-308 y-109}{12 \left(y^2+4
y+1\right)}+\frac{7 y+2}{\left(y^2+4 y+1\right)^2}
  \bigg)
  \text{H}(0,y)
  \nonumber \\
  & & {} 
  +\left(\frac{1}{6} \left(-152+7 \pi ^2\right)+\frac{7 \pi
^2}{6 (y-1)}-\frac{7 \pi ^2}{6 (y+1)}-\frac{24 y}{y^2+4 y+1}\right)
\text{H}(1,y)
  \nonumber \\
  & & {} 
  +\left(\frac{4 (2 y+1)}{y^2+4 y+1}-\frac{62}{3}-\frac{44}{3
(y-1)}-\frac{28}{3 (y+1)}\right)
\text{H}(0,1,y)
  \nonumber \\
  & & {} 
  +\left(\frac{4}{(y-1)^2}+\frac{8}{y+1}+\frac{4}{(y+1)^2}
-8-\frac{8}{y-1}\right) \text{H}(0,-1,0,y)
  \nonumber \\
  & & {} 
  +\left(\frac{3 (2 y+1)}{y^2+4
y+1}+\frac{1}{3}-\frac{43}{6 (y-1)}-\frac{65}{6 (y+1)}\right)
\text{H}(-1,0,y)
	+\left(\frac{2 y}{y^2+4 y+1}\right.
  \nonumber \\
  & & \left. {} 
  -\frac{41}{6}-\frac{31}{3
(y-1)}-\frac{11}{2 (y-1)^2}+\frac{13}{3 (y+1)}+\frac{1}{2 (y+1)^2}\right)
\text{H}(0,0,y)
  \nonumber \\
  & & {} 
  +\left(\frac{-2 y-1}{y^2+4 y+1}-\frac{62}{3}-\frac{13}{6
(y-1)}+\frac{49}{6 (y+1)}\right) \text{H}(1,0,y)
  \nonumber \\
  & & {} 
  +\left(\frac{2}{(y-1)^2}-\frac{22}{y+1}+\frac{2}{(y+1)^2}
+22+\frac{22}{y-1}\right)
\text{H}(0,1,0,y)
  \nonumber \\
  & & {} 
  +\left(-\frac{3}{(y-1)^2}+\frac{4}{y+1}-\frac{3}{(y+1)^2
}-4-\frac{4}{y-1}\right)
\text{H}(0,0,0,y)
  \nonumber \\
  & & {} 
  +\left(
    1
    -\frac{1}{1-y}
    +\frac{1}{(1-y)^2}
    -\frac{1}{1+y}
    +\frac{1}{(1+y)^2}
  \right) 
  \bigg( 
      8 \text{H}(0,0,1,y)
  \nonumber \\
  & & {} 
  + 8 \text{H}(-1,0,0,y)
  - 4 \text{H}(1,0,0,y)
  \bigg)
  -32 \text{H}(1,1,y)
  \nonumber \\
  & & {} 
  +\left(1 - \frac{1}{1-y} -\frac{1}{1+y} \right)
  \bigg( 
    24 \text{H}(-1,0,1,y) 
  +    \text{H}(0,1,1,y)
  -  8 \text{H}(1,0,1,y)
  \nonumber \\
  & & {} 
  + 18 \text{H}(-1,-1,0,y)
  -  6 \text{H}(-1,1,0,y)
  -  6 \text{H}(1,-1,0,y)
  +  2 \text{H}(1,1,0,y)
  \nonumber \\
  & & {} 
  - 21 \zeta(2) \text{H}(-1,,y) 
  - \frac{23}{2} \zeta(3)
  \bigg)
  -\frac{6 (4 y+1)}{\left(y^2+4 y+1\right)^2}
  +\frac{1}{72} \left(272 \pi ^2-1299\right)
  \nonumber \\
  & & {} 
  +\frac{109 \pi ^2}{72 (y+1)}
  +\frac{143 \pi^2 }{72 (y-1)}
  +\frac{-14 \pi ^2 y-276 y-7 \pi^2+72}{12 \left(y^2+4 y+1\right)}
  \Bigg]
  \nonumber \\
  & & {}
  + \alpha^1_{3} \!\left( \mu^2/q^2 ;  y \right)
  + \cO( \epsilon )
  \Bigg\} \,,
 \label{sec::int_antennae::A13}
 \\[2ex]
%
%
  \lefteqn{
  \tilde{\cA}^1_{3} \!\left( \epsilon, \mu^2/q^2 ;  y \right) 
  \; = \;
  %
  \left( \frac{\mu^2}{q^2} \right)^{2 \epsilon}
  \Bigg\{   
  %
  %
  \frac{1}{ \epsilon }
  \Bigg[
  - \left( 
  4
  - \frac{4}{1-y} 
  - \frac{4}{1+y }
  \right)
  \text{H}(1,0,y)
  } \quad
  \nonumber \\
  & & {}
  + \left(
    \frac{15}{12} 
  - \frac{19}{12 (1-y)}
  - \frac{39}{12 (1+y)}
  + \frac{7 (2 y+1)}{3 \left(y^2+4 y+1\right)}  
  \right)
  \text{H}(0,y)
  \nonumber \\
  & & {}
  +\left(
  4
  -\frac{16}{3 (1-y)}
  -\frac{4}{1+y}
  +\frac{4 (2 y+1)}{3 \left(y^2+4 y+1\right)}
  \right) \text{H}(-1,0,y)
  - \left(
  \frac{2 (5 y+1)}{3\left(y^2+4 y+1\right)}
  \right.
  \nonumber \\
  & & \left. {} 
  %
  + 8
  -\frac{37}{6 (1-y)}
  -\frac{7}{2 (1+y)}
  +\frac{3}{2(1-y)^2}
  -\frac{1}{2(1+y)^2}
  \right) \text{H}(0,0,y)
  \nonumber \\
  & & {}
  -\left(
  8
  -\frac{25}{3 (1-y)}
  -\frac{7}{1+y}
  +\frac{4}{(1-y)^2}
  +\frac{4}{(1+y)^2}
  - \frac{2 (2 y+1)}{3 \left(y^2+4 y+1\right)}
  \right)
  \text{H}(0,0,0,y)
  \nonumber \\
  & & {}
  + \left(
  1
  -\frac{1}{1-y}
  -\frac{1}{1+y}
  +\frac{1}{(1-y)^2}
  +\frac{1}{(1+y)^2}
  \right)
  \big( 
  4 \text{H}(0,1,0,y) 
  \nonumber \\
  & & {}
  - 4  \text{H}(0,-1,0 ,y
  + 8 \text{H}(-1,0,0,y)
  - 8 \text{H}(1,0,0,y)
  + 2 \zeta(2) \text{H}( 0, y )
  + 4 \zeta(3)
  \big)
  \nonumber \\
  & & {}
  +\frac{2 \pi ^2 y-72 y+\pi ^2}{9 \left(y^2+4 y+1\right)}
  +\frac{\pi ^2}{3 (1+y)}
  +\frac{2 \pi^2 }{9 (1-y)}
  -\frac{1}{6} \left(2 \pi^2 + 21\right)
  \Bigg]
  \nonumber \\
  & & {}
  + \tilde{\alpha}^1_{3} \!\left( \mu^2/q^2 ;  y \right)
  + \cO( \epsilon )
  \Bigg\}\,.
  \label{sec::int_antennae::A13t}
\end{eqnarray}
The pole parts of the integrated one-loop antennae presented above are given
solely in terms of ordinary harmonic polylogarithms $\text{H}$ as defined
in \cite{Remiddi:1999ew}.  However, the finite remainders $\alpha^1_{3}$
and $\tilde{\alpha}^1_{3}$ of $\cO\! \left( \epsilon^0 \right)$, which we have
also computed analytically, involve cyclotomic polylogarithms. These
expressions are too long to be presented here, but can be obtained from the
authors upon request.

For the sake of completeness, we list the integrated forms of the
remaining two antenna functions $\hat{A}^1_{3,f}$ and $\hat{A}^1_{3,F}$. They
read:
\begin{eqnarray}
  \hat{\cA}^1_{3,f} \!\left( \epsilon, \mu^2/q^2 ;  y \right) & = &
  \frac{1}{3\epsilon} \,
  \Gamma\!\left( 1 + \epsilon \right) e^{\gamma_{\mbox{\tiny E}} \epsilon} \,
  \cA^0_3 \!\left( \epsilon, \mu^2/q^2 ;  y \right) \,,
  \label{sec::int_antennae::A13f}
  \\[2ex]
  \hat{\cA}^1_{3,F} \!\left( \epsilon, \mu^2/q^2 ;  y \right) 
  & = &
  - \frac{1}{3\epsilon} \,
  \Gamma\!\left( 1 + \epsilon \right) e^{\gamma_{\mbox{\tiny E}} \epsilon} \,
  \left[ \left( \frac{ \mu^2 }{ m^2_Q }
  \right)^\epsilon - 1 \right]
  \cA^0_3 \!\left( \epsilon, \mu^2/q^2 ;  y \right) \,,
  \label{sec::int_antennae::A13F}
\end{eqnarray}
where the integrated three-parton tree-level antenna $\cA^0_3$ is given in
\cite{GehrmannDeRidder:2009fz} (see also \cite{Abelof:2011jv}).

An immediate test of whether the IR-divergent pieces (and also the
finite terms) of these antenna
functions are correct is presented in the next section.

 \section{Computation of \boldmath$R_Q$}
\label{sec:compR}

As a first application and, especially, as a check of the IR-singularity
structure as well as of the finite parts of our results for the integrated
antenna functions,  we compute the order  $\alpha_s^2$ contribution to the
inclusive production
 of a massive  quark-antiquark pair by $e^+e^-$ annihilation via a virtual photon
 for arbitrary squared center-of-mass energy above the
                   pair production threshold, $s> 4 m_Q^2$. The
 ratio $R_Q$ is defined by
\begin{eqnarray} \label{RQdef}
    R_Q(s)& = & \frac{ \sigma(e^+\, e^-\, \to \gamma^\ast \to \, Q \bar{Q} + X )}
    {\sigma(e^+\, e^-\, \to \gamma^\ast \to \, \mu^+\,\mu^-) } 
    \nonumber    \\
    & = &  R^{ ( 0 ) }  
    + \left( \frac{\alpha_s ( \mu^2 ) }{ 2 \pi } \right) R^{(1)}
    + \left( \frac{\alpha_s ( \mu^2 ) }{ 2 \pi } \right)^2 R^{(2)}  
    + \cO( \alpha_s^3 ) \, . 
\end{eqnarray}
To order $\alpha_s$, this ratio has  been known for a long time
\cite{Kallen:1955fb,Schwinger:1973rv}. The second order term   $R^{(2)}$   may
be decomposed  into gauge-invariant pieces associated with the different color
structures as follows: 
\begin{equation} \label{RQcostru}
R^{(2)} = e_Q^2 
    \left( N^2_c - 1 \right) \left(
    N_c\, R^{(2)}_{\mss{LC}}
    -\frac{ 1 }{ N_c } \, R^{(2)}_{\mss{SC}}
    + \, n_f\, R^{(2)}_{f}
    + \, R^{(2)}_{F} 
    \right).
\end{equation}
Here, $e_Q$ is the electric charge in units of the positron charge, and
$R^{(2)}_{\mss{LC}}$,  $R^{(2)}_{\mss{SC}}$,  $R^{(2)}_{f}$, and  $R^{(2)}_{F}$
denote the leading color, subleading color, and massless fermion
contributions, and the  heavy-quark contribution from the $Q{\bar Q}$,
   $Q{\bar Q}g$,  and  $Q{\bar Q}Q{\bar Q}$ (above  $s> 16 m_Q^2$) final states, respectively.
 
A complete calculation of  the contribution $R^{(2)}_{f}$ was made first in \cite{Hoang:1995ex}.
Within the antenna framework it was calculated in \cite{Bernreuther:2011jt} in
terms of harmonic polylogarithms and agreement was found between the 
analytical results of \cite{Bernreuther:2011jt} and \cite{Hoang:1995ex}.
Therefore, we do not further consider this term here.  

In \cite{Chetyrkin:1996cf} the leading and subleading color contributions to 
$R^{(2)}$  were computed from the imaginary part of the  order $\alpha_s^2$
photon vacuum
 polarization function $\Pi^{(2)}(q^2)$,  using analytically known
 expansions  of $\Pi^{(2)}$ at $q^2=0$, near the $Q{\bar Q}$ threshold and for
 $-q^2\to\infty$, and  Pad{\'e}  approximations in the range above threshold
 and below the high-energy region. 
 With the same techniques, the   contribution of order  $\alpha_s^3$ 
 to  \eqref{RQdef} was calculated in \cite{Kiyo:2009gb}. Very
 recently, the term  $R^{(2)}$ was computed also in \cite{Gao:2014nva}.

Here we compute $R^{(2)}_{\mss{LC}}$ and  $R^{(2)}_{\mss{SC}}$ by applying
the framework devised in Sec.~\ref{sec:revir} to the case $S=\gamma^*$. Up to a trivial normalization 
$R^{(2)}$ can be obtained by adding up the corresponding
subtracted cross sections \eqref{sbdifcrx}, \eqref{ph3RVfinite}, and \eqref{dsi2nnlosub}
associated with the four-parton, three-parton, and two-parton final states, respectively. 
More precisely, these contributions have to be evaluated with the specific choice 
$J^{(n)}_2 = 1$, which corresponds to inclusive $Q \bar{Q}$ production. 
In this particular case, the subtracted three- and four-parton cross sections
\eqref{sbdifcrx} and \eqref{ph3RVfinite} yield a vanishing
contribution because, by construction, the original matrix elements squared and the subtraction terms
coincide. Therefore, all information on $R^{(2)}$ is encoded in the subtracted
two-parton contribution \eqref{dsi2nnlosub}.
The matrix element which determines the second-oder QCD contribution 
$d\sigma^{VV,Q \bar{Q} }_\mss{NNLO}$ to \eqref{dsi2nnlosub} was computed in
\cite{Bernreuther:2004ih} (and confirmed in \cite{Gluza:2009yy}) and is used
here. The sum of the integrated subtraction terms  in \eqref{dsi2nnlosub} is
given in \eqref{VV::sub}. Adding up the $\epsilon$ poles of $d\sigma^{VV,Q
\bar{Q} }_\mss{NNLO}$ \cite{Bernreuther:2004ih}, of $\cA^0_4, \tilde{\cA}^0_4$
\cite{Bernreuther:2013uma}, of $\cB^0_4$ \cite{Bernreuther:2011jt}, of  
$\cA^0_3$ \cite{GehrmannDeRidder:2009fz}, of   $\cA^1_3$,  
$\tilde{\cA}^1_{3}$,
  $\hat{\cA}^1_{3,f}$, $\hat{\cA}^1_{3,F}$
(cf.~eqs.~\eqref{sec::int_antennae::A13}, \eqref{sec::int_antennae::A13t},
\eqref{sec::int_antennae::A13f}, and \eqref{sec::int_antennae::A13F},
respectively), and of 
 $\delta {\cM}^{\gamma^*}_{(2,1)}$ we find that all IR poles cancel analytically. Therefore,
our result for the NNLO QCD correction \eqref{RQcostru} to $R_Q$ is
indeed finite.

In addition to \eqref{dsi2nnlosub}, the contribution from the $Q{\bar
Q}Q{\bar Q}$ final state has to be taken into account, which has not been
discussed in Sec.~\ref{sec:revir}.
The four-particle phase-space integral of the squared tree-level
matrix element of  $\gamma^*\to Q{\bar Q}Q{\bar Q}$ is completely finite and
can be evaluated numerically. (The respective phase-space integrals cannot be
expressed solely by polylogarithms.)
Denoting this contribution to  \eqref{RQdef} by  $R^{(2)}_{4Q}$, its color
structure is
\begin{equation} \label{cosrr2F}
  R^{(2)}_{4Q} = e_Q^2 \left( N^2_c - 1 \right) 
     \left[R^{(2)}_{F,4Q}  - \frac{1}{N_c}  R^{(2)}_{E} \right] \, ,
\end{equation}
 where $R^{(2)}_{E}$ is an interference term which results from the
 fact that there are two identical
 (anti)quarks in the final state. This term makes a contribution to the subleading
 color term $R^{(2)}_{\mss{SC}}$ in  \eqref{RQcostru}.
 It becomes logarithmically divergent in the  high-energy limit
 $m_Q^2/s\to 0$, (cf.~\cite{Catani:1999nf}):
 \begin{equation} \label{RcontEtr}
 R^{(2)}_{E} =  
 \left(\frac{13}{8}-\frac{\pi^2}{4} + 
   \zeta(3)\right)\ln(s/m_Q^2) + \frac{c}{4}  + {\cal
O}\left(\frac{m_Q^2}{s}\right) ,
\end{equation}
 where $\zeta(x)$ denotes the Riemann zeta function. 
  In \cite{Catani:1999nf} the constant $c$ was extracted from
 a numerical computation of $ R^{(2)}_{E} $ in the region of small $m_Q^2/s$ and 
 found to be $c = -8.7190 \pm 0.0013$.
 In the limit $m_Q^2 / s \to 0$, the contribution from
\eqref{dsi2nnlosub} to $R^{(2)}_{\mss{SC}}$
   exhibits
  the same logarithm as in  \eqref{RcontEtr} but with
  opposite sign. As a result,  $R^{(2)}_{\mss{SC}}$ remains finite
  for $m_Q\to 0$ as it should be. For the constant $c$ in
  \eqref{RcontEtr} we obtain
\begin{equation}
 c = -45 +\frac{5 \pi ^2}{2}
 -\frac{19 \pi ^4}{90}
 +4 \pi^2 \ln (2)
 +4 \zeta (3)
 \approx - 8.7175\,,
\end{equation}
which is in good agreement with the numerical value of Ref.~\cite{Catani:1999nf}
cited above.
  
\begin{figure}
  \centering
  \includegraphics[width=.8\linewidth]{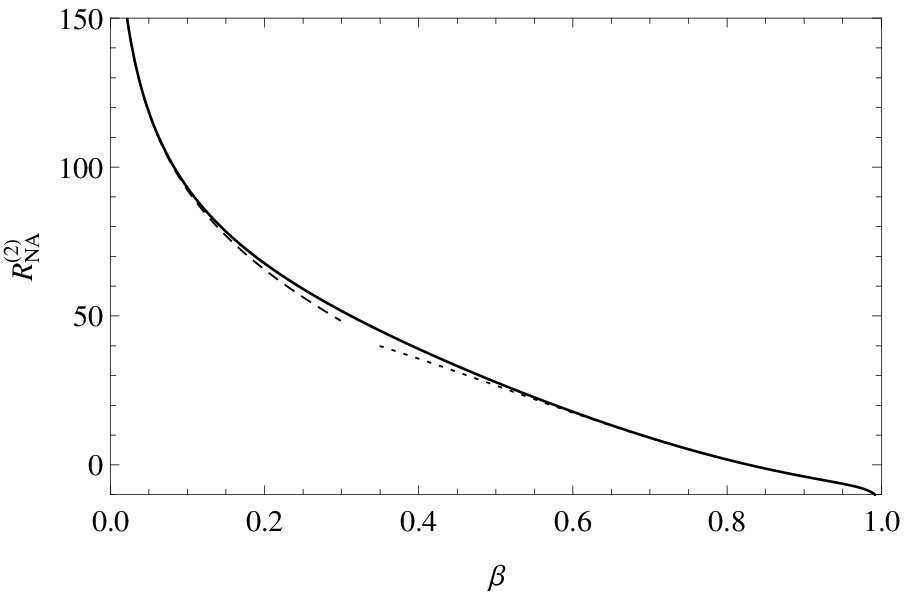} \\[2ex]
  \includegraphics[width=.8\linewidth]{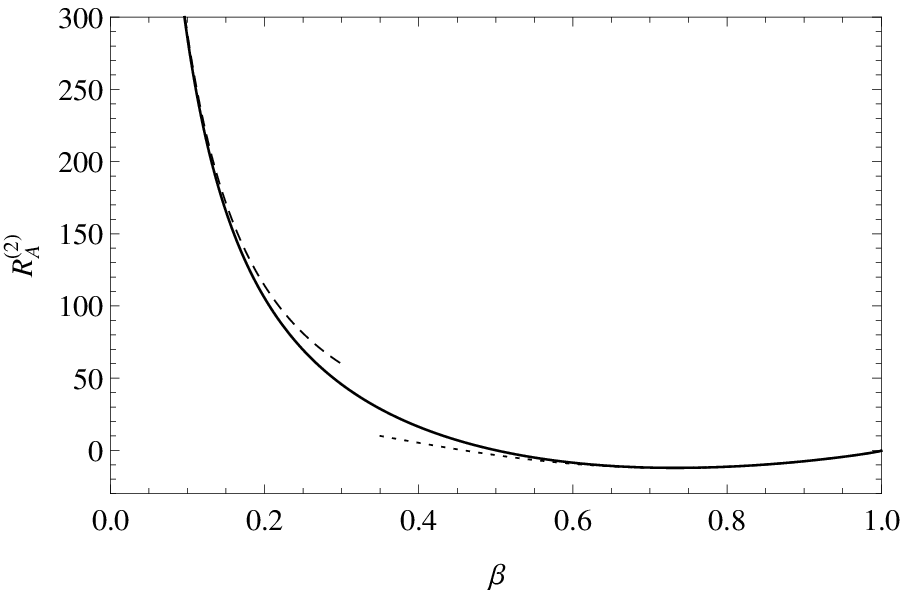}
  \caption{Our exact results for $R^{(2)}_{\mss{NA}}$ and
  $R^{(2)}_{\mss{A}}$ plotted against $\beta$ (solid line). 
  The renormalization scale is chosen to be $\mu = m_Q$.
  For comparison, the expansions at the kinematic threshold (dashed
  curves) \cite{Hoang:1997sj,Czarnecki:1997vz} and in the asymptotic region
  (dotted curves) \cite{Chetyrkin:1997qi} are included as
  well.}
  \label{fig:RNARA}
\end{figure}

In order to compare our results for  the leading and subleading color
contributions with the results and the asymptotic expansions given 
in\cite{Gorishnii:1986pz,Chetyrkin:1990kr, 
Chetyrkin:1994ex,Chetyrkin:1997qi,Chetyrkin:1996cf} and with the threshold
expansion of \cite{Hoang:1997sj,Beneke:1997jm,Czarnecki:1997vz}, we switch to
the color decomposition used in  these papers and define
\begin{equation} \label{RnewColor}
 R^{(2)}_{\mss{NA}} = \frac{N_c}{2}\left( R^{(2)}_{\mss{LC}}
 - R^{(2)}_{\mss{SC}} \right) \, ,
 \qquad  R^{(2)}_{\mss{A}} = N_c  R^{(2)}_{\mss{SC}} \, .
\end{equation}
The solid lines in Fig.~\ref{fig:RNARA} show our results  for
$R^{(2)}_{\mss{NA}}$ and  $R^{(2)}_{\mss{A}}$ in the range
 $0 < \beta \leq 1$. These results are exact -- the usual harmonic
 polylogarithms 
 which enter the expressions for $R^{(2)}_{\mss{NA}}$ and $R^{(2)}_{\mss{A}}$
 are evaluated numerically with the \texttt{Mathematica} package \texttt{HPL}
\cite{Maitre:2005uu}, whereas the remaining cyclotomic polylogarithms are
evaluated by utilizing their defining integral representations
\cite{Ablinger:2011te}.
 For $\beta\to 0$, the behavior of 
 $R^{(2)}_{\mss{A}}$ is dominated by the Coulomb singularity $\sim
 1/\beta$, while $R^{(2)}_{\mss{NA}}$ diverges only logarithmically
 for $\beta\to 0$.

In the  high-energy limit  $x=m_Q^2/s\to 0$  the terms
$R^{(2)}_{\mss{NA/A}}$ are known as asymptotic expansions in $x$.
In \cite{Chetyrkin:1997qi} these expansions are  given up to  and including
terms of order $x^6$. These asymptotic expressions are shown by the dotted curves in 
Fig.~\ref{fig:RNARA}. We have expanded our exact results for
$R^{(2)}_{\mss{NA/A}}$ around $x=0$ with the help of the \texttt{Mathematica}
packages \texttt{HPL} \cite{Maitre:2005uu} and  \texttt{HarmonicSums}
\cite{Ablinger:2011te,Ablinger:2013cf} and compared with the expansions given
in \cite{Chetyrkin:1997qi} and find agreement.  In case
of the leading color contribution $R^{(2)}_{\mss{LC}}$ this comparison is
performed in fully analytic fashion. Because we have not computed
   the higher-order terms in the
expansion \eqref{RcontEtr} in analytical fashion, we checked the small-$x$ 
behavior of $R^{(2)}_{\mss{SC}}$ only numerically.

In the literature, $R^{(2)}_{\mss{NA}}$ and $R^{(2)}_{\mss{A}}$  are known
analytically near the  $Q{\bar Q}$ production threshold as an expansion in
 $\beta$ up to and including terms of order $\beta$
 \cite{Czarnecki:1997vz} (cf. also
 \cite{Beneke:1997jm,Hoang:1997sj}). 
 These expansions are shown as dashed  lines in
 Fig.~\ref{fig:RNARA}. To this order in $\beta$ both the non-abelian and
 abelian term  $R^{(2)}_{\mss{NA/A}}$  is determined by the 
 second order QCD matrix element of $\gamma^*\to Q{\bar Q}$. As stated
 above we use this matrix element from  \cite{Bernreuther:2004ih}. The
 threshold expansion of this  matrix element  agrees with 
\cite{Czarnecki:1997vz}.
  We checked by small-$\beta$ expansion of the polylogarithms 
 which enter our expressions for 
 \eqref{dsi2nnlosub}
 that the three- and four-parton final states 
 contribute  to
$R^{(2)}_{\mss{NA/A}}$  only with higher powers in $\beta$. 
 Explicit expansions of $R^{(2)}_{\mss{NA/A}}$
 to higher orders in  $\beta$ will be presented elsewhere \cite{DABB2014}.

Ref.~\cite{Chetyrkin:1996cf} computed $R^{(2)}_{\mss{NA/A}}$ in the
 whole physical region  $0 < \beta \leq 1$ by using the asymptotic and
 the small $q^2$ expansions  of the photon vacuum
 polarization function $\Pi^{(2)}(q^2)$ as input for Pad{\'e}
 approximation  of $\Pi^{(2)}(q^2)$ (whose imaginary part yields
 $R^{(2)}$) away from the threshold and the asymptotic region.
 Figs.~5 of \cite{Chetyrkin:1996cf} show the result of this
 calculation, in particular  $R^{(2)}_{\mss{NA}}$ and 
 $R^{(2)}_{\mss{A}}$ as functions of $\beta$. The comparison of these
 results with our results of  Fig.~\ref{fig:RNARA}  shows 
 very small differences only in the interval  $0.15\lesssim \beta\lesssim
 0.5$. For instance , 
 $|\Delta R^{(2)}_{\mss{NA}}/ R^{(2)}_{\mss{NA}}|, |\Delta
 R^{(2)}_{\mss{A}}/ R^{(2)}_{\mss{A}}|\lesssim 1\%$
 at $\beta=0.3$.
   
Thus  our antenna functions pass this
non-trivial test.

\section{Summary and Outlook}
\label{sec:concl}

We have computed the integrated real-virtual antenna functions 
 which are required, within the antenna subtraction framework,  for
 the calculation of the  differential  cross section for massive quark-pair production
 by an uncolored initial state at NNLO QCD. This completes, together
 with our previous results
 \cite{Bernreuther:2011jt,Bernreuther:2013uma} on the subtraction terms for the matrix
 elements of the final states $Q{\bar Q} q{\bar q}$ and  $Q{\bar Q} g
 g$,  the set of antenna subtraction terms which are required for this type of
 processes. As a check of our result, we have computed the inclusive
 cross section, respectively the ratio $R_Q(s)$ for $e^+e^-\to\gamma^*\to
 Q{\bar Q} X$ at order $\alpha_s^2$ by adding up the
 antenna-subtracted  cross sections associated with the  two-parton, three-parton, and
 four-parton final states. Our result is infrared-finite as it
 should, and our second order correction to   $R_Q(s)$, which we
 obtained without any approximation, agrees with the  known threshold
 \cite{Czarnecki:1997vz} and
 asymptotic \cite{Chetyrkin:1997qi} expansions in the respective
 kinematical limits.

Future applications of these subtraction terms include  the
computation of differential distributions for $S\to Q{\bar Q} X$ at
NNLO QCD. Our (integrated) antenna
functions are, in addition, also important ingredients for the
antenna-subtraction approach to NNLO QCD analyses  of  hadro-production of massive quark pairs.


\subsubsection*{Acknowledgements}

  We wish to thank T. Gehrmann for helpful discussions and
  J. Ablinger and J. Bl\"umlein for information on the expansion of
  cyclotomic harmonic polylogarithms. 
The work of O.D. is supported by the Cluster of Excellence Precision Physics,
Fundamental Interactions and Structure of Matter (PRISMA – EXC 1098).
The work of  W.B. is supported by Deutsche Forschungsgemeinschaft, SFB/TR9.


\end{document}